\documentclass[a4paper,11pt]{article}
\pdfoutput=1 

\usepackage{jcappub} 

\usepackage[T1]{fontenc} 
\usepackage{makecell}

\usepackage{lineno}

\usepackage{xcolor}
\usepackage[normalem]{ulem}
\usepackage{soul}

\usepackage{ulem}
\usepackage{subcaption}

\usepackage{appendix}

\title{\boldmath Model of the instrumental polarization induced by a rotating HWP with anti-reflective sub-wavelength structures for CMB polarimetry}

\author[a,1]{R. Takaku,\note{Corresponding author.}}
\author[b]{H. Imada}
\author[c]{G. Patanchon}
\author[d]{Y. Sakurai}

\affiliation[a]{Okayama University, 3-1-1 Tsushima-naka, Kita-ku, Okayama 700-8530, Japan}
\affiliation[b]{National Astronomical Observatory of Japan, 2-21-1 Osawa, Mitaka, Tokyo 181-8588, Japan}
\affiliation[c]{Universit\'e Paris-Cit\'e, CNRS, Astroparticule et Cosmologie, Paris F-75013, France}
\affiliation[d]{Suwa University of Science, 5000-1 Toyohira, Chino, Nagano 391-0292, Japan}

\emailAdd{ryota.takaku@okayama-u.ac.jp}
\emailAdd{guillaume.patanchon@apc.univ-paris-diderot.fr}
\emailAdd{}
\emailAdd{sakurai\_yuki@rs.sus.ac.jp}

\abstract{

Future CMB polarization experiments require precise control of instrumental systematics in the search for primordial $B$-mode polarization. A rotating half-wave plate (HWP) provides effective polarization modulation, but its non-ideal electromagnetic response can generate HWP-synchronous instrumental polarization. In this work, we characterize the response of a multi-layer sapphire HWP with anti-reflective sub-wavelength structures (SWS) using rigorous coupled-wave analysis. For three asymmetric SWS models, we find no substantial differences in the transmittance, polarization efficiency, or instrumental-polarization response that can be attributed to the type of SWS asymmetry. The fourth-harmonic instrumental polarization relevant to CMB polarization reconstruction is typically of order $10^{-4}$ in the CMB bands, which is expected to be mitigated. We further evaluate an actually fabricated SWS geometry, whose asymmetry enhances the retardance, and compare its response with that of an idealized model having the same type of asymmetry. Comparison of the realistic SWS with the idealized model shows that higher transmittance tends to be associated with lower instrumental polarization. These results suggest that improving the total transmission, including the effect of the fabricated SWS profile, may be more important for reducing HWP-induced instrumental polarization than tuning the SWS asymmetry itself.
}

\begin{document}
\maketitle
\flushbottom
\section{Introduction}

Measurement of the cosmic microwave background (CMB) temperature and polarization fluctuations allows us to constrain the physics of the early universe.
The Planck~\cite{Planck1} satellite mission has measured the CMB anisotropies with unprecedented accuracy, leading to a measurement of cosmological parameters with percent precision, and opened the possibility to test the theory of inflation~\cite{Planck:2018jri}, a phase of rapid expansion of the universe, which is a hypothesis to explain problems of the standard cosmological model such as the flatness problem, the horizon problem and the origin of structures. 
Two types of fluctuations were generated during inflation, scalar (density) and tensor (gravitational wave) modes, and they both induce polarization of the CMB after Compton interaction with local anisotropies during the last scattering of photons with electrons.
The primordial gravitational wave leaves a distinct divergence-free polarization pattern on the sky, called primordial B-mode. One of the main objectives of the cosmology community is currently to detect and measure this tiny signal which would provide a direct measurement of the primordial gravitational wave signal, and it would give access to the physics of inflation.
This signal has not yet been detected, and many experiments are targeting it~\cite{ACT:2020gnv,BICEP:2021xfz,Essinger-Hileman:2014pja,GroundBIRD:2020wax,LB_PTEP,Li:2017drr,LSPE:2020uos,POLARBEAR:2022dxa,QUBIC:2020kvy,SimonsObservatory:2018koc,SPT:2019nip}.

The major challenges for this measurement come from two aspects. 
The first one is the need for broadband observations of the sky emission from a few tens of gigahertz to terahertz, because of the presence of polarized galactic thermal dust and synchrotron emissions that are added to the CMB along the line of sight, even at high galactic latitudes.
We can separate the various galactic polarized emissions from the CMB by using the difference in the emission spectra: galactic dust has an increasing spectrum while synchrotron has a decreasing spectrum relative to the CMB. 

The second challenge is the potential systematic effects associated with polarimetry which requires differencing measurements, and those are potentially taken under different conditions. Mismatches can originate from system instabilities in time, and from gain and beam shape differences between detectors with orthogonal polarizers.
Those mismatches make a conversion from intensity to polarization when one detector output is subtracted from the orthogonal one.
The primordial B-mode signal is at the level of nano-Kelvin and is added to the uniform CMB blackbody radiation field at 3\,K, the dipolar signal at 3.36\,mK~\cite{Planck1}, and the CMB temperature fluctuations of the order of $100~\mu$K.
Moreover, the detector instability could lead to large 1/f noise, which is propagated to the polarization signal if data are combined (and differenced) at very different time scales with different detectors to separate the polarization signal from the intensity.
Thus, for an accurate CMB polarization observation one has to minimize any instrumentally induced polarization systematic effect over all the frequency bands.

\begin{figure}
    \centering
    \includegraphics[width=0.7\linewidth]{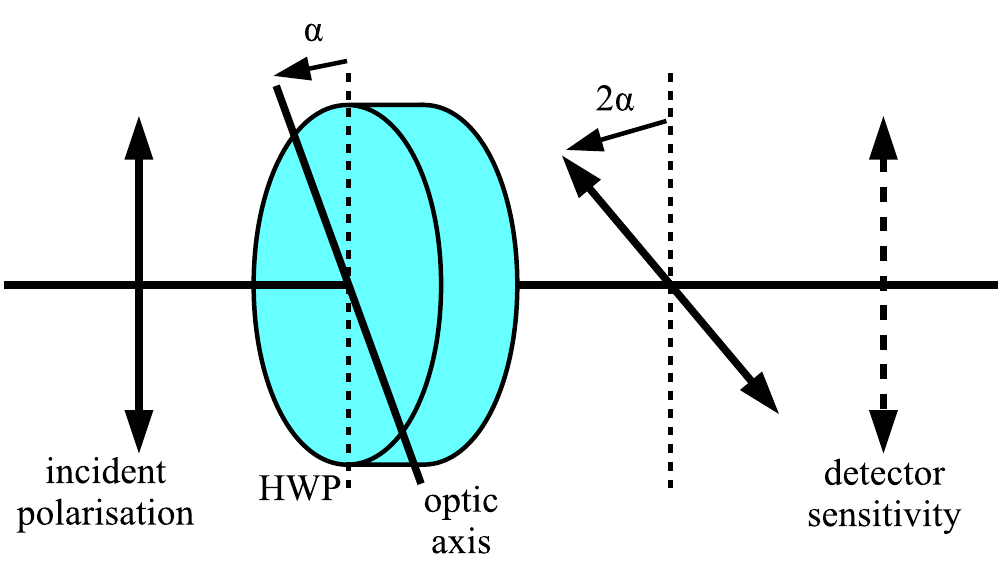}
    \caption{A schematic diagram of a polarization modulator based on an HWP. The HWP rotates the incident linear polarization by $2\alpha$, where $\alpha$ is defined as the angle between the optic axis and incident polarization.}
    \label{fig:modulation_system}
\end{figure}

To overcome these systematic errors, a number of ongoing and design-stage CMB polarization telescopes include a polarization modulator using a rotating half-wave plate(HWP)~\cite{Kusaka_2018,ACTPol2016,Takakura_2017,Salatino2018, Sullivan2018,komatsu2019,Buzzelli2017}.
An ideal HWP induces a rotation to an incident electric field by $2\alpha$, where $\alpha$ is the angle between one of the axis tangential to the HWP~\cite{hecht2021optics} and the incident field, as shown in Figure~\ref{fig:modulation_system}.
When the HWP rotates, the outgoing linearly polarized signal measured by a detector also rotates four times faster, with an amplitude equal to that of the incident polarization.
Therefore, a rotating HWP works as a linear polarization modulator.
A rapidly rotating HWP is particularly attractive because it allows us to reconstruct Stokes Q and U quasi-instantaneously using only a single detector by making several measurements per rotation, avoiding mismatches arising when using different detectors.
Fast modulation shifts the signal to higher frequencies, where we expect white noise. This drastically reduces the leakage of $1/f$ noise into the estimated polarization maps.

While there are benefits in using HWP, the HWP itself can introduce specific systematic effects.
The main identified effect is the conversion of intensity to polarization, hereafter referred to as instrumental polarization, especially at four times higher frequency of the HWP rotation~\cite{Patanchon_2024}.
HWP non-idealities can also induce mixing of Q and U components and reduce the efficiency of polarization detection.
Several studies have been carried out in the CMB community to characterize HWP imperfections in the context of ground experiments~\cite{Bryan:10, Essinger-Hileman:13, Essinger-Hileman:14}.
Bryan et al.~\cite{Bryan:10} discussed the effect of a non-ideal HWP and their correction assuming a normal incidence and source spectra for the band-averaged Mueller matrix.
ABS~\cite{Kusaka_2018} has employed a HWP and Kusaka et al.\cite{Kusaka2014} has addressed the potential instrumental polarization by a single HWP.
Recently Simons Observatory (SO) is observing with cryogenic rotating HWPs and their performances are reported \cite{Sugiyama2024, Yamada2024, Yamada2026}.

There are several reports on systematic estimation of HWP non-idealities for the satellite, e.g. LiteBIRD~\cite{LB_PTEP}.
Giardiello et al.~\cite{Serena2022} gives a model HWP imperfections and estimates how these imperfections propagate into the estimation of the cosmological parameter.
Monelli et al. \cite{Monelli_2024} adopt frequency-dependent Mueller matrices derived from realistic HWP designs, and investigate the effect on cosmological estimation in cleaning foreground.
Stellati et al.~\cite{Stellati_2026} focuses on the pointing errors induced by HWP imperfection, especially the misalignment of the rotation axis of HWP and a wedge effect. Micheli et al.~\cite{Silvia} studies the impact of TES non-linearity coupled to HWP imperfections on cosmological parameter estimation, particularly through signals at twice the HWP rotation frequency. 
Patanchon et al. \cite{Patanchon_2024} investigate the instrumental polarization caused by the non-ideal response of a rotating HWP, quantify its impact on the reconstructed CMB B-mode signal, and propose a maximum-likelihood correction method in the map-making.
From the instrumentation, there are several reports on characterization of the HWP~\cite{Pisano:2014eba,Komatsu_9layer,Takaku2023}, and show imperfections by the measurements.

In principle a single layer of HWP is not used for a HWP over a broadband coverage, because the retardance in the HWP depends on frequency.
We achieve a broadband performances when we stack multi-layer HWP stacking with specific relative optic axes~\cite{Pancharatnam1955}.
Several kinds of design for millimeter wavelength range have been developed~\cite{Komatsu_2020}.
The major material for the HWP is sapphire because it has high-thermal conductivity, low dielectric loss at millimeter wavelength, and large differential refractive index.
However, since the refractive index of sapphire is more than 3.0, it requires a broadband anti-reflection coating (ARC) on the HWP surfaces.
A typical way of ARC is multiple thin layer coatings having different refractive indices~\cite{2018SPIE10708E..43N}.
But it has a potential risk of delamination at a cryogenic operation because of different thermal contractions between layers.
Recently, applying sub-wavelength structures (SWS) gets attractive as a broadband ARC especially at a cryogenic operation.
It achieves a broadband ARC with the optical element itself when the physical shape of SWS is scaled to the required observation band, thus it avoids the risk of delamination.
This has been applied for the actual optical elements, e.g. lens, filter, and window for cosmological and astrophysical observations~\cite{2013ApOpt..52.8747D,2022ApOpt..61.8904G,2026ApOpt..65.4579T,Takaku:21}.
ARC performance directly affects to the level of HWP imperfection, thus it is important to consider the design of ARC together with HWP design itself.

This paper aims at constructing a detailed effect of a transmissive multi-layer HWP with SWS on the electromagnetic field,
for the study of the impact of related systematic effect, particularly focusing on the instrumental polarization, which was shown to be a major issue for B-mode measurements at large angular scales~\cite{Patanchon_2024}.
Such a characterization is important for future broadband CMB polarization experiments, as SWSs provide a promising approach to reducing reflection from HWP surfaces over a broad frequency range. 
In this work, we use a previous LiteBIRD configuration as a realistic test case for a broadband multi-layer sapphire HWP~\cite{LB_PTEP}. The results are also relevant to other CMB polarimeters employing sapphire-based rotating HWPs.
We provide a methodology for the calculation of Mueller matrix coefficients based on rigorous coupled-wave analysis (RCWA)~\cite{Moharam:95} method, which solves the Maxwell equation of periodic gratings in the harmonic space with the periodic boundary condition, with respect to the observation frequency, and incident angle and the azimuthal angle. For this last dependence, we focus on a few relevant harmonics of the HWP rotation frequency.
In Section~\ref{sec:model}, we define the coordinate systems and angles to specify the polarization states as a Stokes vector.
We introduce the Mueller calculus used to model the HWP, which relate the incident angle to the detector output.
The electromagnetic simulation model of the HWP with SWS is illustrated in Section~\ref{sec:simulation}. Here we provide three types of SWS which make different retardance combining the birefringence of the material itself.
We show the outcome of the EM simulation in Section~\ref{sec:results}.
We focus on the relevant parameters converting intensity to polarization.
Based on the simulation results, we assess which type of SWS can better mitigate instrumental polarization in Section~\ref{sec:discussion}, and we compare the results with coefficients obtained with actual fabricated structures.

\section{Model}
\label{sec:model}
Let us first introduce the system configuration and the coordinate systems and the relevant angles to describe the polarization states on the sky, an HWP, and linearly polarized detectors. 
We adopt the Mueller formalism to express the properties of the HWP as a function of the rotation angles, and write a general form of the Mueller matrix and the output of a linearly polarized detector.

\subsection{System}
\label{subsec:system}
In this paper, we assume that the system is composed of a HWP; a telescope without any aberrations; and a detector sensitive to linear polarization, arranged in the same ordering. we do not model diffraction at the edge of the HWP and telescope aperture.

The HWP is a stack of parallel flat plates made of sapphire, a uniaxial birefringent material characterized by ordinary and extraordinary refractive indices. The extraordinary axis, or optic axis, is parallel to the surface of the HWP.
It is equipped with SWS.
The details of the SWS geometry are provided in Section~\ref{sec:simulation}.

\subsection{Definitions of coordinate systems}
\label{subsec:coordination}
\begin{figure}
    \centering
    \begin{subfigure}[b]{0.32\linewidth}
        \centering
        \includegraphics[width=\linewidth]{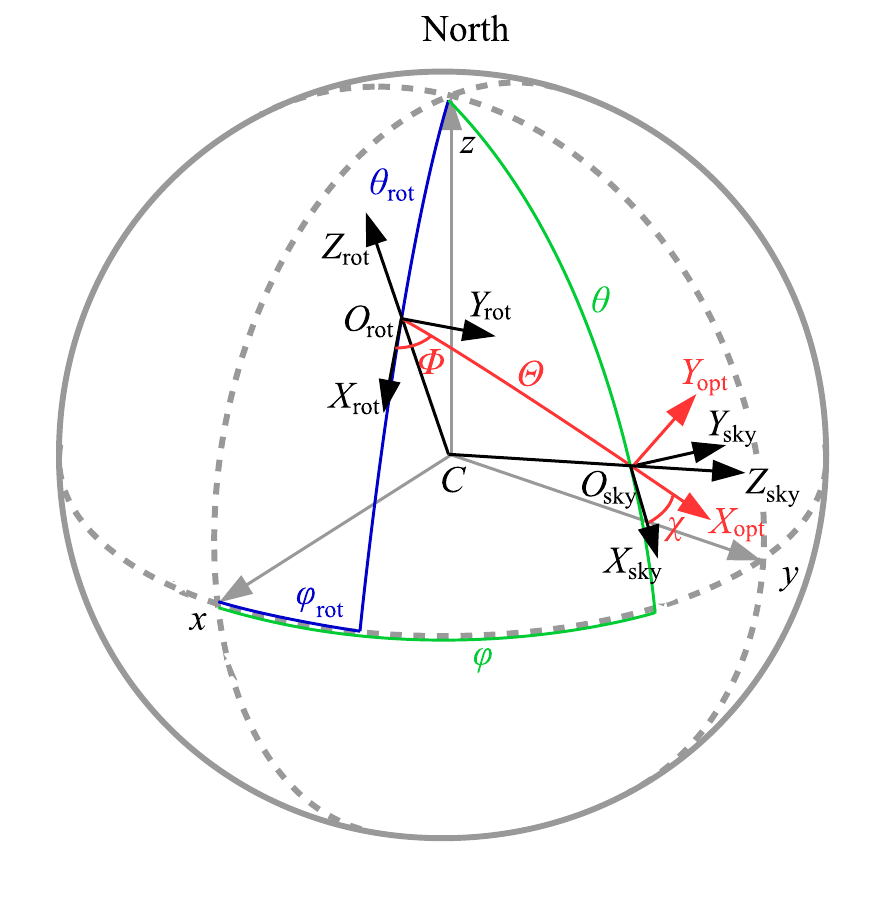}
        \caption{}
        \label{fig:coordination}
    \end{subfigure}
    \hfill
    \begin{subfigure}[b]{0.64\linewidth}
        \centering
        \includegraphics[width=\linewidth]{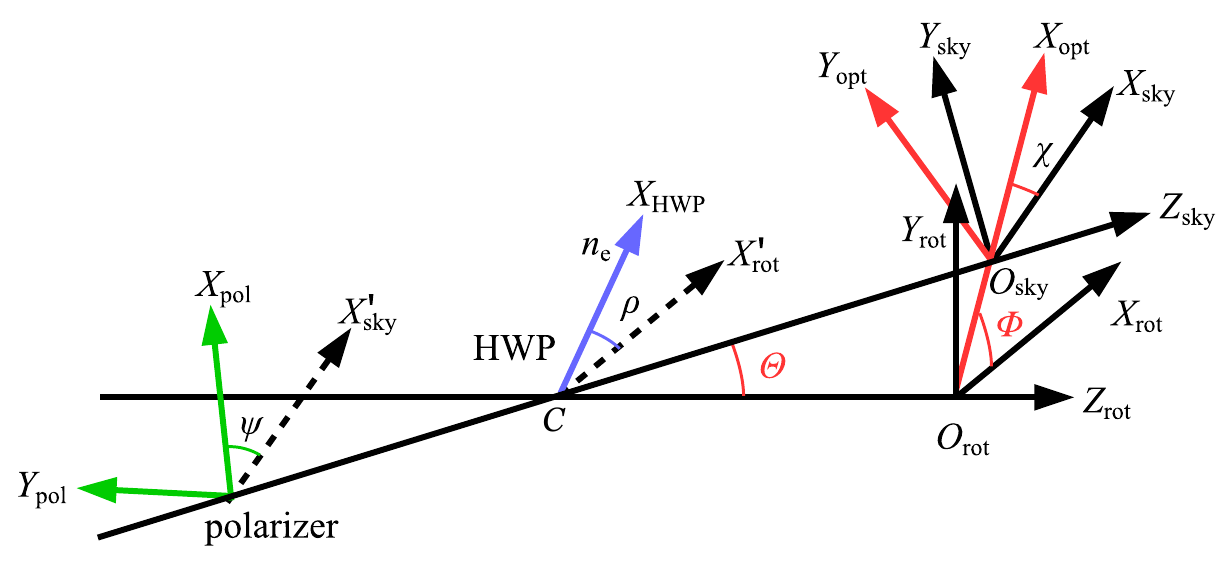}
        \caption{}
        \label{fig:axis_relation}
    \end{subfigure}
    \caption{
        Definition of (a) the coordinate systems and
        (b) the angles and axes.
        Both panels share the triangle
        $O_{\mathrm{rot}}CO_{\mathrm{sky}}$.
    }
    \label{fig:coordination_all}
\end{figure}

Suppose the celestial sphere in Figure~\ref{fig:coordination} and the telescope at the center of the sphere (at the position $C$).
We introduce the spherical coordinates $(\theta,\phi)$ of the direction of the observed incoming plane wave, as shown in Figure~\ref{fig:coordination} in green.
The coordinates $(\theta_{\mathrm{rot}},\phi_{\mathrm{rot}})$ in blue denote the direction toward which the rotation axis of an HWP points on the celestial sphere.
The scanning strategy of CMB observation determines the angles $(\theta_{\mathrm{rot}},\phi_{\mathrm{rot}})$ and $(\theta,\phi)$.
The $X_{\mathrm{rot}}$ and $Y_{\mathrm{rot}}$ axes in Figure~\ref{fig:coordination} are along the lines of isolongitude and isolatitude, respectively.
For simplicity, we assume that the rotation axis $\overrightarrow{CO_{\mathrm{rot}}}$ is normal to the surface of the HWP.
Two angles $\Theta$ and $\Phi$ in red are the incident and azimuthal angle of the incoming wave with respect to the HWP, respectively.
The incident angle $\Theta$ is then the angular distance between the two points $(\theta, \phi)$ and $(\theta_{\mathrm{rot}},\phi_{\mathrm{rot}})$.
The angle $\Phi$ is defined with respect to $X_{\mathrm{rot}}$.

According to the convention used in HEALPix~\cite{healpix}, which is often employed for CMB analysis, a purely positive Stokes $Q$ is defined such that the electric field is linearly polarized parallel to the $X_{\mathrm{sky}}$ axis, and a purely positive Stokes $U$ is defined as a linear polarization rotated by $45$ degrees counterclockwise seen from the positive $Z_{\mathrm{sky}}$. 
Stokes $V_{\mathrm{sky}}$ of the incoming wave is defined with the basis $\{(\hat{e}_{X,\mathrm{sky}} \pm \sqrt{-1}\hat{e}_{Y,\mathrm{sky}})/\sqrt{2}\}$ for positive and negative $V$, where $\hat{e}_{X,\mathrm{sky}}$ and $\hat{e}_{Y,\mathrm{sky}}$ represent unit vectors parallel to $X_{\mathrm{sky}}$- and $Y_{\mathrm{sky}}$-axes, respectively. 
Although the CMB is not expected to be circularly polarized, we leave this possibility in particular for other astrophysical sources. 
This is important since imperfections of the HWP can induce conversion from circular to linear polarization~\cite{Eimer2024, Nagy2017, Fujino2025}. For the optical calculations, $p$- and $s$-polarizations, which are defined with respect to the plane of incidence~\cite{hecht2021optics}, are often used except for normal incidence. 
The Stokes parameters $Q_{\mathrm{opt}}$ and $U_{\mathrm{opt}}$ of the incoming wave are defined in the optical coordinate system defined by the $X_{\mathrm{opt}}$ and $Y_{\mathrm{opt}}$ axes, using the same sign and angle convention as the HEALPix convention described above.
$X_{\mathrm{opt}}$ axis is along the great circle passing through the two points $O_{\mathrm{rot}}$ and $O_{\mathrm{sky}}$. 
A purely positive $Q_{\mathrm{opt}}$ corresponds to $p$-polarization and a purely negative $Q_{\mathrm{opt}}$ to $s$-polarization. 
Here we introduce the focal lane orientation in the sky $\chi$ as the angle between $X_{\mathrm{sky}}$ and $X_{\mathrm{opt}}$ axes, which is also a function of $(\theta_{\mathrm{rot}}, \varphi_{\mathrm{rot}})$ and $(\theta,\varphi)$.

The rotation angle of the HWP, $\rho$, is defined as the angle made by the optic axis of the HWP and the projection of $X_{\mathrm{rot}}$ axis onto the HWP in Figure~\ref{fig:axis_relation}. 
We finally introduce the orientation of the detector polarizer $\psi$ with respect to the projection of $X_{\mathrm{sky}}$ axis onto the plane of the detector. 
The relations between various angles are summarized in Figure~\ref{fig:axis_relation}.



\subsection{Mueller matrix and incident signal to detector}
\label{subsec:mueller_matrix}

We define the Stokes vector in the sky coordinate system as
$\mathbf{S}_{\mathrm{sky}} = (I_{\mathrm{sky}}, Q_{\mathrm{sky}}, U_{\mathrm{sky}}, V_{\mathrm{sky}})$.
We also define the Stokes vector in the optical coordinate system as
$\mathbf{S}_{\mathrm{opt}} = (I_{\mathrm{opt}}, Q_{\mathrm{opt}}, U_{\mathrm{opt}}, V_{\mathrm{opt}})$.
They are related by $\mathbf{S}_{\mathrm{opt}} = R(\chi)\mathbf{S}_{\mathrm{sky}}$, where $R$ is a rotation matrix defined as
\begin{align}
R(\chi)
&=
\begin{pmatrix}
1 & 0 & 0 & 0 \\
0 & \cos 2\chi & \sin 2\chi & 0 \\
0 & -\sin 2\chi & \cos 2\chi & 0 \\
0 & 0 & 0 & 1
\end{pmatrix}.
\end{align}

We use the Mueller-matrix formalism to describe the propagation of the Stokes vector through the HWP and the detector polarizer~\cite{hecht2021optics}.
The intrinsic HWP properties depend on the incident angle $\Theta$ and the relative azimuthal angle $\rho-\Phi$.
We therefore describe the HWP by the Mueller matrix $M(\Theta,\rho-\Phi)$ in the telescope and focal plane.
The detector is sensitive to a single linear polarization direction, and its Mueller matrix is described as

\begin{align}
G
&=
\frac{1}{2}
\begin{pmatrix}
1 & 1 & 0 & 0 \\
1 & 1 & 0 & 0 \\
0 & 0 & 0 & 0 \\
0 & 0 & 0 & 0
\end{pmatrix}.
\end{align}
By combining the previous quantities, the power ($S$) incident on a linearly polarized detector is given by
\begin{align}
\label{eq:Signal}
2S
&= (1,\ 0,\ 0,\ 0)\, G R(\psi) R(-\chi) T M(\Theta,\rho-\Phi) R(\chi)\mathbf{S}_{\mathrm{sky}} \nonumber \\
&= \bigl(1,\ \cos 2(\psi-\chi),\ \sin 2(\psi-\chi),\ 0\bigr)\,
   \times T M(\Theta,\rho-\Phi) R(\chi)\mathbf{S}_{\mathrm{sky}},
\end{align}
where ($\psi-\chi$) represents the orientation of the detector polarizer with respect to the ($X_{\mathrm{opt}}$) axis, and $T$ denotes the Mueller matrix of the telescope, which is assumed to be the identity matrix.

In the case of an ideal HWP, shifting the phase of one linear polarization by $\pi$, the corresponding Mueller matrix $M_{\mathrm{idl}}$ is given by
\begin{align}
\label{eq:ideal_mueller}
M_{\mathrm{idl}}(\Theta,\rho-\Phi)
&=
\begin{pmatrix}
1 & 0 & 0 & 0 \\
0 & \cos 4(\rho-\Phi) & \sin 4(\rho-\Phi) & 0 \\
0 & \sin 4(\rho-\Phi) & -\cos 4(\rho-\Phi) & 0 \\
0 & 0 & 0 & -1
\end{pmatrix},
\end{align}
and substituting Eq.~\ref{eq:ideal_mueller} into Eq.~\ref{eq:Signal}, we obtain
\begin{align}
\label{eq:2Sidl}
2S_{\mathrm{idl}}
&=
I_{\mathrm{sky}}
+ Q_{\mathrm{sky}} \cos\left(4\rho - 4\Phi + 4\chi - 2\psi\right) \notag \\
&\quad
+ U_{\mathrm{sky}} \sin\left(4\rho - 4\Phi + 4\chi - 2\psi\right).
\end{align}


For a non-ideal HWP, $M$ is not as simple as in Eq.~\ref{eq:ideal_mueller}. 
The calculation of this Mueller matrix is described in the next section.
Non-idealities in the Mueller-matrix coefficients are expected even for an HWP fabricated exactly as designed, because a realistic HWP has finite reflection, frequency-dependent retardance, and differential transmission.
These effects introduce non-zero elements $M_{IX}$ and $M_{XI}$ ($X = Q,U,V$), corresponding to instrumental polarization.
Their impacts on systematics are particularly important when they contain components varying with the azimuthal angle at four times the HWP rotation frequency, i.e. proportional to a combination of $\cos(4\rho-4\Phi)$ and $\sin(4\rho-4\Phi)$. 
Such components convert temperature anisotropy into a spurious linear polarization with the same modulation as the incident sky polarization and cannot be separated from it by the standard demodulation~\cite{Patanchon_2024}.
Harmonics other than the $4f_{\mathrm{HWP}}$ component do not directly project onto the recovered sky-polarized signal.

\section{Simulation}
\label{sec:simulation}

\subsection{HWP modeling}
\label{subsec:HWP_modeling}

\begin{figure}
    \centering
    \includegraphics[width=0.5\linewidth]{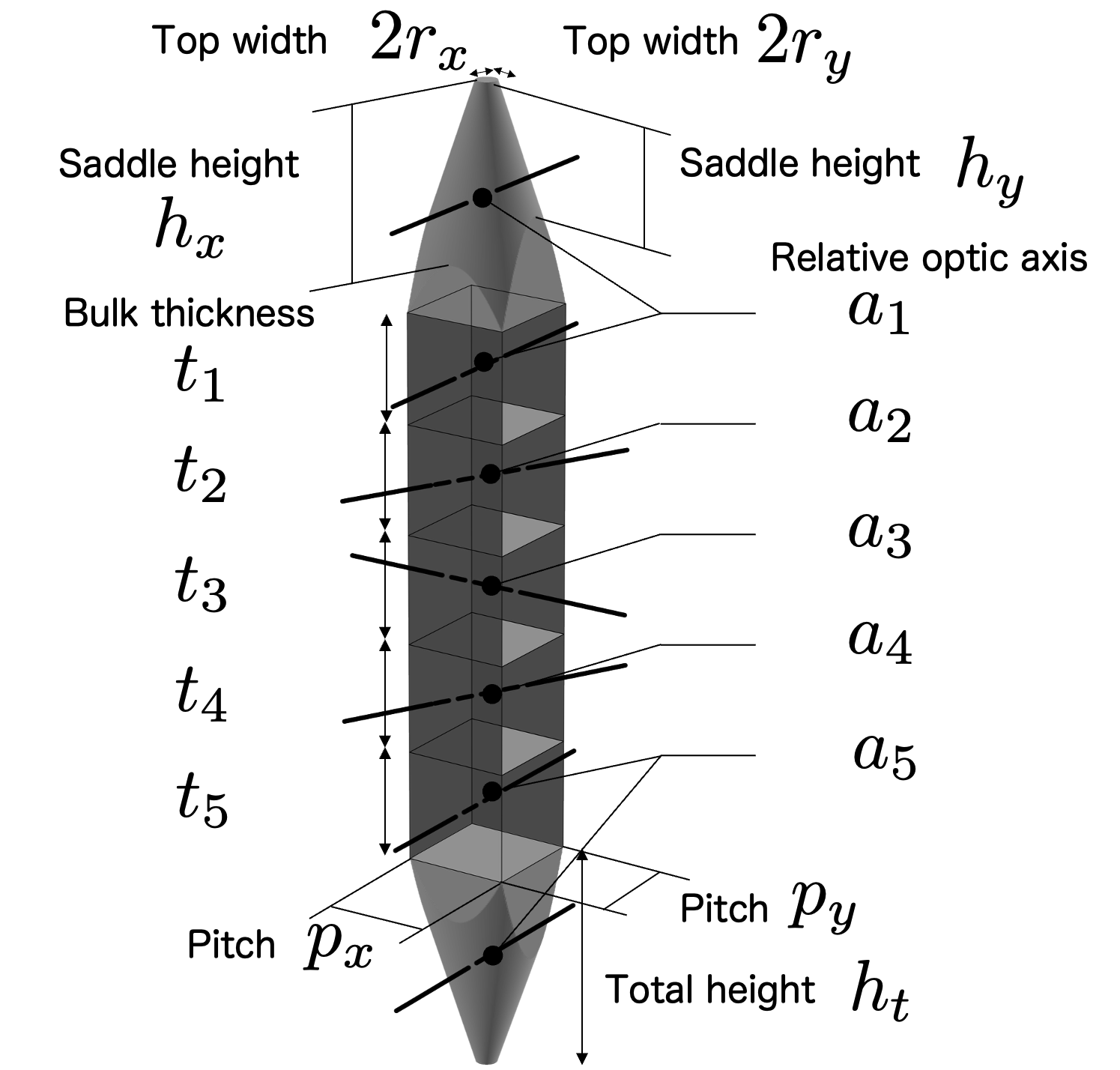}
    \caption{Unit cell of HWP model for RCWA simulation. The gray part represents sapphire. The structure is equipped for AR, and its size is not in scale.}
    \label{fig:rcwa_settings}
\end{figure}


We constructed a HWP model for electromagnetic simulation with an SWS, as shown in Figure~\ref{fig:rcwa_settings}.
The assumed HWP is composed of five birefringent flat plates. 
Plates 2--4 have a typical thickness of 4.9 mm, while asymmetric pyramidal SWS are formed on the external faces of plates 1 and 5, defined in Sec.~\ref{subsec:sws}. 
Multi-layer HWP stacked with different relative optic axis gives a broadband achromatic HWP. 
The relative optic-axis angles are listed in Table~\ref{tab:rcwa_settings} to cover the frequency bands defined in Table~\ref{tab:LB_bandwidth}, which configuration is proposed by Komatsu et al.~\cite{Komatsu_2020}.
We used the symmetric model; the angle of the optic axis is symmetric along the 3rd plate. 
The angle of the optical axis of structure is the same as the 1st and 5th plates because the structures are fabricated directly on the outer surfaces of the plates.

\begin{table}[htbp]
    \centering
    \caption{Parameters for RCWA simulations}
    \label{tab:rcwa_settings}
    \begin{tabular}{c|c|c|c|c}
        Parameter & Variable & Model-A & Model-B & Model-C \\\hline\hline
        \multicolumn{5}{c}{Structure parameters}\\\hline\hline
        Pitch $x$           & $p_x$ [mm]     & \multicolumn{3}{c}{0.48} \\\hline
        Pitch $y$           & $p_y$ [mm]     & \multicolumn{3}{c}{0.48} \\\hline
        Top radius $x$      & $r_{xt}$ [mm]  & \multicolumn{3}{c}{0.05} \\\hline
        Top radius $y$      & $r_{yt}$ [mm]  & \multicolumn{3}{c}{0.05} \\\hline
        Total height        & $h_t$ [mm]     & \multicolumn{3}{c}{2.0} \\\hline
        Bottom radius $x$   & $r_{xb}$ [mm]  & 0.315 & 0.390 & 0.45 \\\hline
        Bottom radius $y$   & $r_{yb}$ [mm]  & 0.411 & 0.326 & 0.301\\\hline
        Saddle height $x$   & $h_x$ [mm]     & 1.44 & 1.12 & 0.95 \\\hline
        Saddle height $y$   & $h_y$ [mm]     & 1.19 & 1.50 & 1.63 \\\hline\hline
        \multicolumn{5}{c}{HWP parameters}\\\hline\hline
        Optic axis \#1       & $a_1$ [deg]    & \multicolumn{3}{c}{0}\\ \hline
        Optic axis \#2       & $a_2$ [deg]    & \multicolumn{3}{c}{-26.97}\\ \hline
        Optic axis \#3       & $a_3$ [deg]    & \multicolumn{3}{c}{-88.65}\\ \hline
        Optic axis \#4       & $a_4$ [deg]    & \multicolumn{3}{c}{-26.97}\\ \hline
        Optic axis \#5       & $a_5$ [deg]    & \multicolumn{3}{c}{0}\\ \hline
        Internal plate thickness & $t_{2,3,4}$ [mm] & \multicolumn{3}{c}{4.9} \\\hline
        External plate thickness & $t_{1,5}$ [mm] & 2.9 & 4.9 & 5.8 \\
    \end{tabular}
\end{table}

\begin{table}[htbp]
    \centering
    \caption{Central frequencies and fractional bandwidths of the frequency bands considered in this work. Values are taken from \cite{LB_PTEP}.}
    \label{tab:LB_bandwidth}
    \begin{tabular}{c||c|c|c|c|c|c|c|c|c}
        Frequency [GHz] & 40 &50 &60 &68 &78 &89 &100 &119 &140   \\\hline
        Fractional bandwidth & 0.3 &0.3 &0.23 &0.23 &0.23 &0.23 &0.23 &0.3 &0.3  
    \end{tabular}
\end{table}

\begin{figure}
    \centering
    \includegraphics[width=1.0\linewidth]{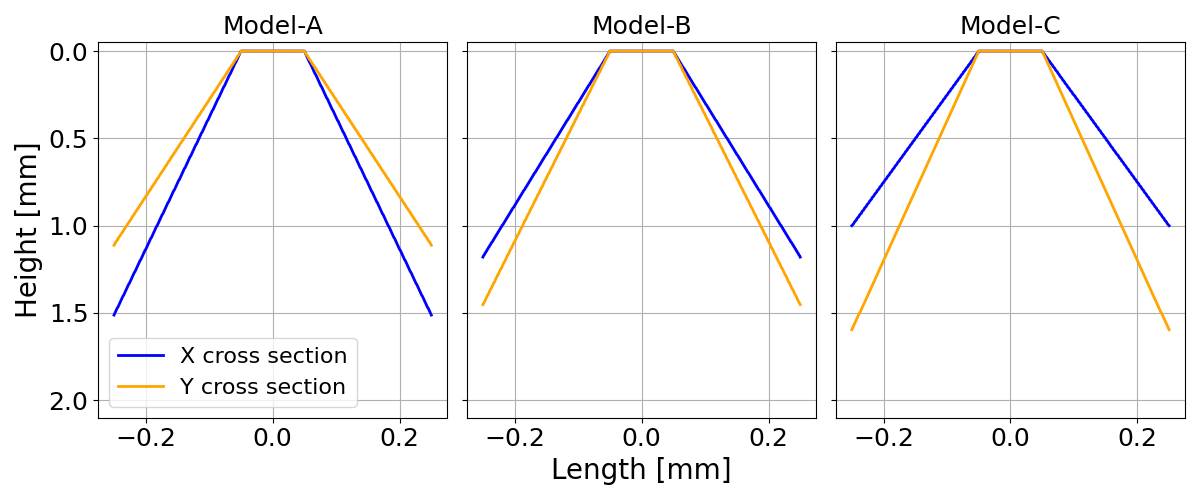}
    \caption{Cross-section view of three-types of asymmetric SWS.}
    \label{fig:cross_section}
\end{figure}

\begin{figure}
    \centering
    \includegraphics[width=0.8\linewidth]{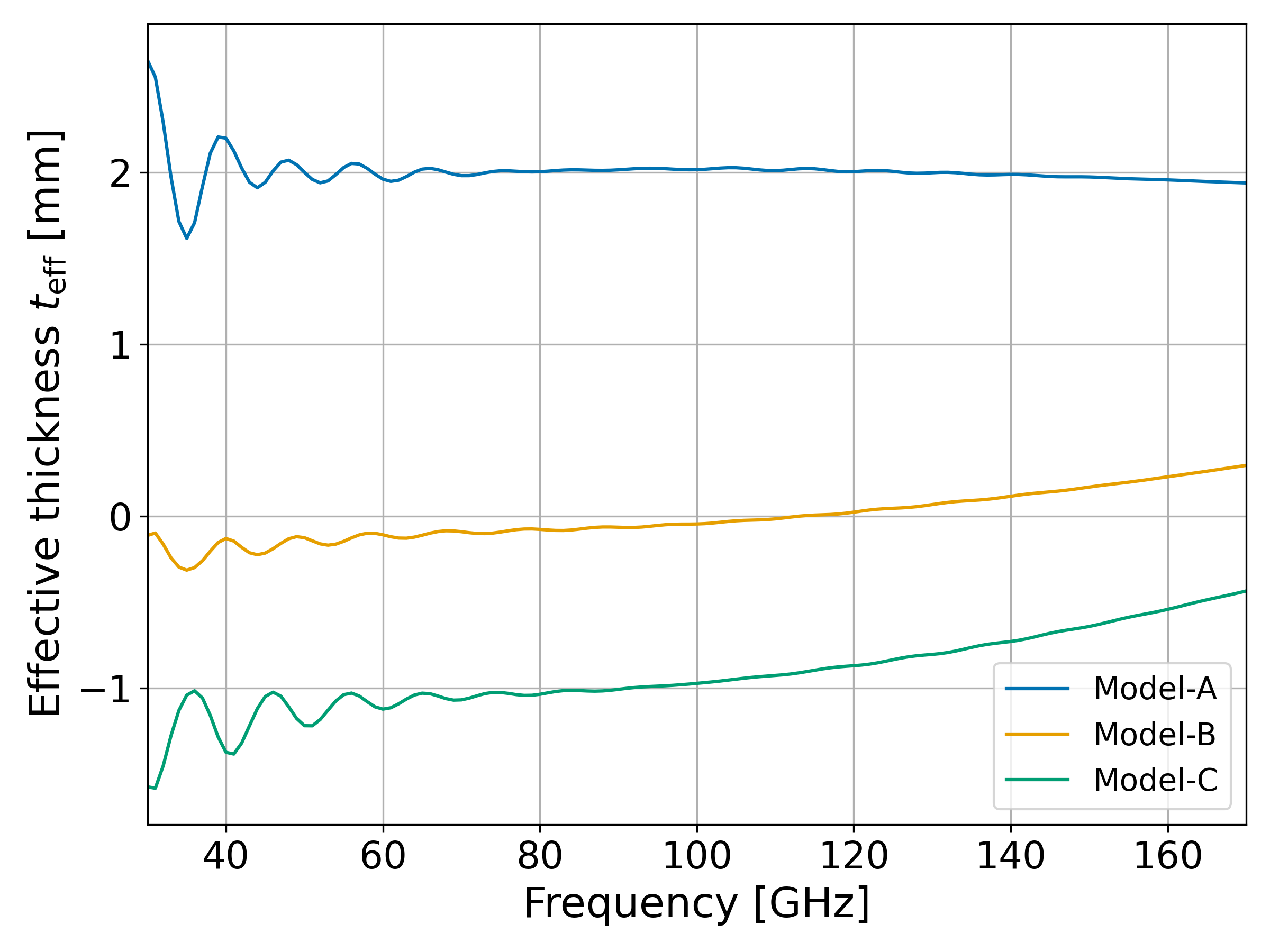}
    \caption{Effective thickness of three types of asymmetric structure}
    \label{fig:teff}
\end{figure}

\subsection{Sub-wavelength structures}
\label{subsec:sws}

SWS on sapphire are commonly fabricated by ultra-short pulsed laser ablation, and their profiles can be described by a geometrical model based on an elliptical contour, especially at the bottom of structures~\cite{Takaku2020,Takaku2023}.
In this model, the radius of the elliptical cross section in the $x$ and $y$, i.e. $r_{xb},r_{yb}$ directions increase with depth.
When the elliptical contour exceeds the unit-cell boundary, it is clipped at the boundary, producing saddle heights in the $x$ and $y$ directions in addition to the total SWS height.
Changing the bottom radii modifies these saddle heights and therefore controls the structural asymmetry and the resulting retardance.
The detail modeling of this SWS are described in Appendix \ref{appendix:SWS}.

We selected three representative asymmetric SWS models to investigate whether different types of SWS asymmetry affect the level of instrumental polarization. Since the asymmetry also changes the additional phase retardance, we adjusted the bulk thicknesses of the outer plates accordingly.
We define geometrical parameters to cover the LFT bands. We define $p_x = p_y = p = 0.5$ mm to avoid higher-order diffraction at higher frequency of LFT. The total height is defined as 2 mm, and top radius is defined as $r_{xt} = r_{yt} = r_t = 0.05$ mm.
We define three representative shapes as shown in Figure~\ref{fig:cross_section}. 
An asymmetric SWS produce an additional retardance because its effective refractive indices differ along the two principal directions. 
This contribution is combined with the phase retardance originating from the birefringence of sapphire. 
Since the effective refractive indices of the structure depend on frequency~\cite{brawer}, its contribution to the retardance also depends on frequency. We express this contribution in terms of a frequency-dependent effective thickness~\cite{Takaku_SPIE2022} and use it to adjust the bulk thickness required to satisfy the half-wave condition.

We used the commercial software DiffractMOD~\cite{DiffractMOD} based on the algorithm: rigorous coupled-wave analysis (RCWA) \cite{Moharam:95}, to calculate the components of Jones matrix through an AR structure.
We calculate the effective thickness of SWS by equations described in R. Takaku et al.~\cite{Takaku_SPIE2022}, as shown in Figure~\ref{fig:teff}.
We assume the refractive indices of birefringent material for ordinary and extraordinary rays to be $n_o = 3.047$ and $n_e = 3.361$, respectively, which are close to the values at the liquid helium temperature \cite{Johnson2004_MAXIPOL}.

The structure of Model-A has the effective thickness of 2~mm, thus both physical and effective thickness are the same. The effective thickness of Model-B averaged in the given frequency band is -0.023 mm, which effectively cancels the retardance introduced by the birefringence in the structure. While one of Model-C is -0.93 mm, which requires the additional retardance in the bulk. We take into account the thickness adjustment to model the broadband achromatic HWP, which is described in the next subsection.
In addition to these idealized SWS models, we also evaluate a realistic SWS geometry based on a fabricated structure and compare its optical response with Model-A in Section~\ref{subsec:realistic_case}.

\subsection{Simulation settings and outputs}
\label{subsec:sim_outputs}
We evaluated the angle dependence of the Mueller matrix with the incident angle from 0 to 10 degrees every 1 degree, at the central frequency of 89 GHz.
We also calculated the propagation of EM wave from 30 to 165 GHz spaced by 0.5 GHz, 
with fixed incident angle at 10 degrees. 
This frequency resolution is sufficient to cover a band-averaged Mueller matrix with a relative width of about 20\,\%. 
To capture the HWP properties at relevant harmonics of the rotation frequency especially at $4(\rho - \Phi)$ 
, we calculate the Mueller matrix at $\rho = 15n$ degrees $(n\in \mathbb{N})$. 

The incident plane wave is assumed to be linearly polarized.
The accuracy of the simulation was of the order of $10^{-6}$ at most. We estimate our results are accurate down to the level of $< 10^{-5}$. This significance is common to all of our simulation results.

The simulation is conducted with DiffractMOD and it calculates the component of Jones matrices. Given an incident angle and frequency, the Mueller matrix is calculated as
\begin{align}
    M_{ij, \Theta,\nu, }(\rho) = \frac{1}{2}\mathrm{Tr}(\sigma_{i}J_{\Theta,\nu}(\rho)\sigma_{j}J_{\Theta,\nu}(\rho)^{\dagger}), ij \in \{0,1,2,3\}.
\end{align}

We obtain $4\times 4$ components of the Mueller matrix as a function of rotation angle. For each component of Mueller matrix, we fit the data with the following functional form:

\begin{align}
M_{XY,\mathrm{fit}}(\rho)
&=
M_{XY, \Theta,\nu}^{(0)}
+ \sum_{m}
\left[
B_{XY, \Theta,\nu}^{(2m)} \sin(2m\rho) + C_{XY, \Theta,\nu}^{(2m)} \cos(2m\rho)
\right], ~m \in \{1,2,3,4\},
\label{eq:reconstruction_sc_en}
\end{align}
This decomposition is useful because we expect only a few harmonics to contribute to the total Mueller matrix.

We obtain the amplitudes and phases of the modulated signal for the different harmonics with
\begin{align}
    M_{XY, \Theta,\nu}^{(2m)} &= \sqrt{ \left(B_{XY, \Theta,\nu}^{(2m)}\right)^2 + \left(C_{XY, \Theta,\nu}^{(2m)}\right)^2} \nonumber\\
    2m\phi^{(2m)}_{XY, \Theta,\nu} & = \operatorname{atan2}(C_{XY, \Theta,\nu}^{(2m)},B_{XY, \Theta,\nu}^{(2m)}).
\end{align}
We then calculate the band-averaged quantities for a sky component $I_s(\nu)$ as
\begin{align}
\mathcal{M}_{XY}^{(2m)} &=
\frac{
\int_{\nu_1}^{\nu_2} d\nu\,
\tau(\nu)\,
I_s(\nu)\,
M_{XY}^{(2m)}(\nu)
\exp\!\left(\sqrt{-1}\phi_{XY}^{(2m)}(\nu)\right)
}{
\int_{\nu_1}^{\nu_2} d\nu\,
\tau(\nu)\,
 I_s(\nu)
},
\nonumber\\
\varepsilon_{XY}^{(2m)} &= \left|\mathcal{M}_{XY}^{(2m)}\right|, \qquad
\varphi_{XY}^{(2m)} =
\arg \mathcal{M}_{XY}^{(2m)} - 2i\,(\rho-\Phi),
\label{eq:bandavg_complex_mueller}
\end{align}
where $\tau(\nu)$ is the band-pass transmission. 
In this paper, we assume a $\tau(\nu)=1$ for $\nu_1 \leq \nu \leq \nu_2$ and zero otherwise.
For CMB temperature anisotropies, $I_s(\nu)$ corresponds to $\partial B(\nu,T_{\rm CMB})/\partial T$ ($T_{\rm CMB} = 2.7255$~K), while for foreground components it is given by the corresponding emission spectrum.
The frequencies $\nu_1$ and $\nu_2$ are the edge of the each frequency band, as shown in Table~\ref{tab:LB_bandwidth}.
From this band average, we obtain the transmittance $\mathcal{M}_{II}^{(0)}$, and the instrumental polarization of fourth harmonics such as $\mathcal{M}_{IX}^{(4)}$ and $\mathcal{M}_{X I}^{(4)}~(X = Q, U)$.
We define another variable quantifying the modulation performance without the effect of the imperfect transmittance, i.e., the ratio of the polarization fraction of the modulated linear polarization to the outgoing energy. 
We refer to it as  polarization efficiency, and it is calculated as $\mathcal{E}_L = \mathcal{M}_{XY}^{(4)}/\mathcal{M}_{II}^{(0)}~(X,Y = Q,U)$.

\section{Results}
\label{sec:results}
We first summarize the harmonic components of the Mueller matrix across the frequency bands considered in this work. 
The instrumental polarization is dominated by the second and fourth harmonics, with amplitudes typically ranging from $10^{-4}$ to $10^{-2}$ and from $10^{-6}$ to $10^{-3}$, respectively. 
Higher-order harmonics are significantly smaller, generally below $10^{-5}$. 
Since the $4f$ harmonics directly contributes to the reconstructed CMB polarization, we focus primarily on this component in the following analysis. 
We show the detail results in this section.

\subsection{Dependence on incident angle}
\label{subsec:sim_inc_angle}

\begin{figure}
    \centering
    \includegraphics[width=1.0\linewidth]{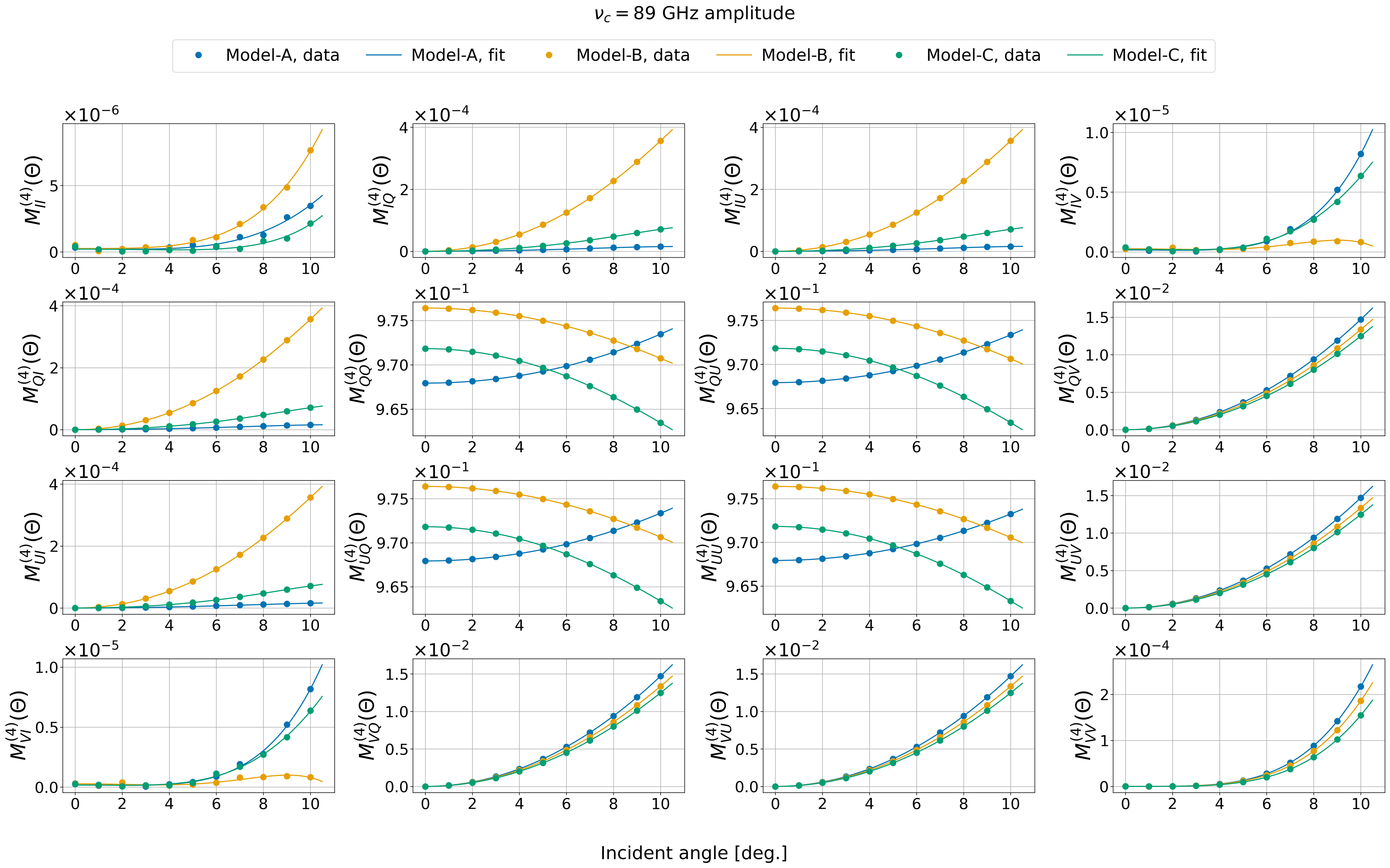}
    \caption{Amplitude $M_{XY}^{(4)}$ as a function of the incident angle $\Theta$ at 89~GHz. The dot and crossed points show the simulation result of the amplitude and phase, respectively. The blue, orange and green colored data show the results for Model- A, B and C, respectively. The curves are obtained by the fitting Equation~\ref{eq:fit_inc_angle}. }
    \label{fig:Mueller_inc_angle_dependence_one_freq}
\end{figure}

\begin{figure}
    \centering
    \includegraphics[width=1.0\linewidth]{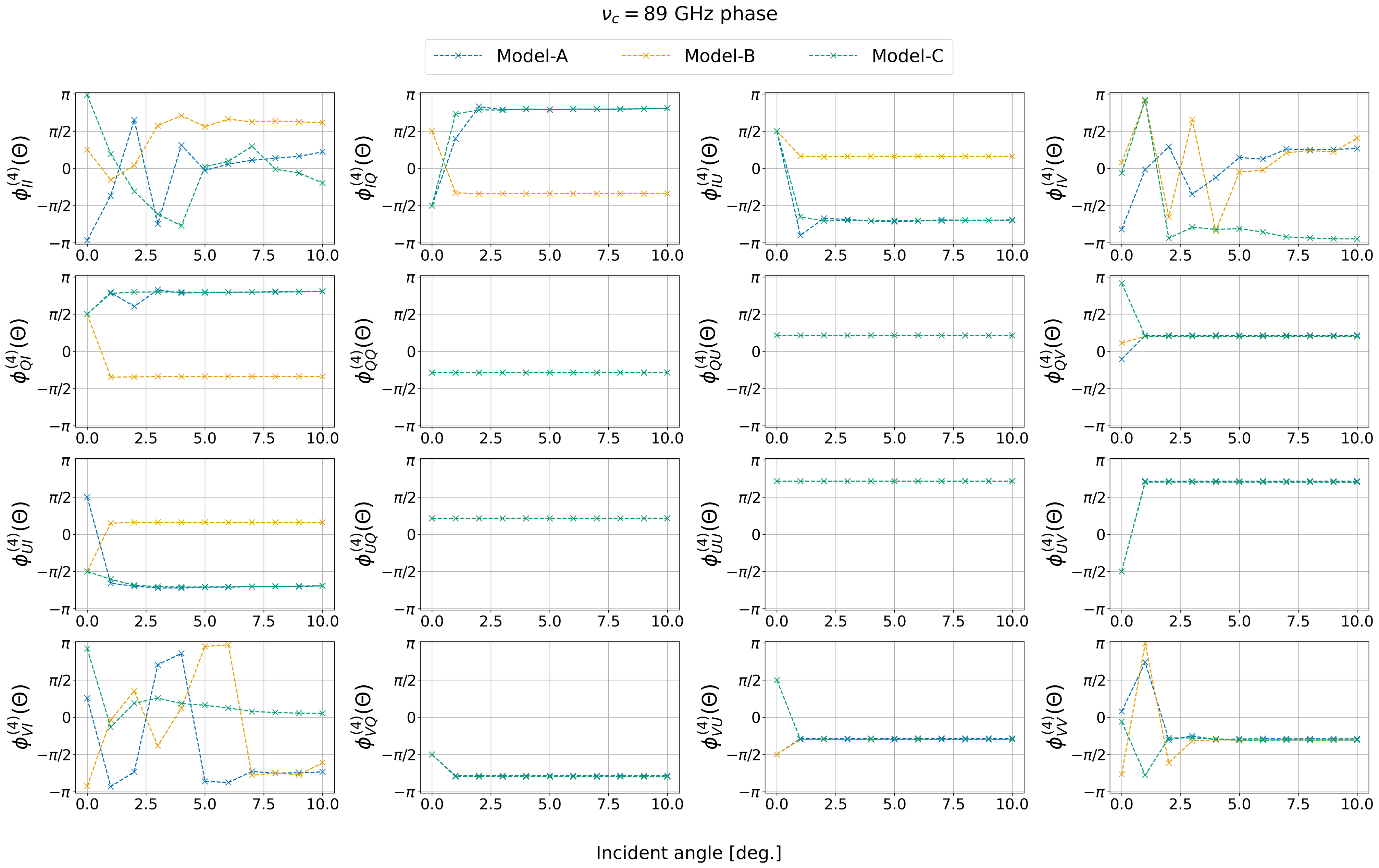}
    \caption{The same as Fig.~\ref{fig:Mueller_inc_angle_dependence_one_freq} for the phase $\phi_{XY}^{(4)}$ as a function of the incident angle $\Theta$ at 89~GHz.}
    \label{fig:Mueller_phase_inc_angle_dependence_one_freq}
\end{figure}

Figure ~\ref{fig:Mueller_inc_angle_dependence_one_freq} and \ref{fig:Mueller_phase_inc_angle_dependence_one_freq} show the results of $M_{XY}^{(4)}$ and $\phi_{XY}^{(4)}$ at 89~GHz as a function of the incident angle, and the comparison for three different SWS models. The phase is independent of the incident angle $\Theta$ up to the precision of simulations. However, as expected, the amplitudes depend strongly on the incident angle. We use the following empirical form to describe the simulated angular dependence:
\begin{align}
\label{eq:fit_inc_angle}
    M_{XY}^{(2m)} (\Theta) = A + B\Theta^2 + C\Theta^4 + D\Theta^6
\end{align}
The coefficients for all $M_{XY}^{(4)}$ are derived by this fitting and summarized in Table ~\ref{tab:Mxy_fit} for $2m = 4$. 

As Equation~\ref{eq:2Sidl} indicates, the linear polarization signal from the sky is modulated up to a frequency of $4(\rho-\Phi)$.
Let us focus on the instrumental polarization at a frequency of $4\rho$ in particular, $M_{IX}^{(4)}(\Theta)$ and $M_{XI}^{(4)}(\Theta)$ ($X = Q,U$), which corresponds to the conversion between un-polarized light and linear polarization. 
This is the major systematics because of large intensity $I$ of the CMB.
As already stated, the phase is independent of the incident angle up to 10 degrees.
At normal incidence, the instrumental polarization does not appear, 
however, when $\Theta = 10^{\circ}$, it reaches the level of $10^{-3}$, especially at the lowest frequency band. In the CMB channels, the amplitude is typically of the order of $10^{-4}$. 
This corresponds to a $4f$ signal of $10^{-4}\times 3~\rm K$ from the CMB monopole and $10^{-7}\times 3~\rm K$ from the dipole in the TOD.

Another type of conversion, $M_{VX}^{(4)}(\Theta)$ and $M_{XV}^{(4)}(\Theta)$ ($X = Q,U$) corresponding to the conversion between circular and linear polarization, gets at the level of $10^{-2}$ for almost all of frequency bands and models. 
Ref.~\cite{CircularPol} predicts intrinsic circular polarization from galactic synchrotron emission at a level of $\delta V \sim 10^{-8}\,{\rm K}$ at 10 GHz on degree angular scales.
The corresponding leakage into linear polarization is below $0.1\,{\rm nK}$ and is negligible.

\begin{table*}[t]
\centering
\small
\setlength{\tabcolsep}{4pt}
\caption{Fitting results of $M_{XY}^{(2)}(\Theta)$ at 89 GHz using $a+b x^2+c x^4+d x^6$, where $x=\Theta/10^\circ$. In each parameter block, columns A, B, and C correspond to Model-A, Model-B, and Model-C, respectively.}
\label{tab:Mxy_fit}
\begin{tabular}{c|ccc|ccc}
\hline
& \multicolumn{3}{c|}{$a$} & \multicolumn{3}{c}{$b$} \\
$XY$ & Model-A & Model-B & Model-C & Model-A & Model-B & Model-C \\
\hline
$II$ & $2.2\times 10^{-7}$ & $2.55\times 10^{-7}$ & $2.12\times 10^{-7}$ & $-4.44\times 10^{-7}$ & $1.75\times 10^{-7}$ & $-5.45\times 10^{-7}$ \\
$IQ$ & $1.63\times 10^{-7}$ & $8.24\times 10^{-8}$ & $3.2\times 10^{-8}$ & $1.84\times 10^{-5}$ & $3.32\times 10^{-4}$ & $6.44\times 10^{-5}$ \\
$IU$ & $8.58\times 10^{-8}$ & $1.23\times 10^{-7}$ & $-8.93\times 10^{-8}$ & $1.82\times 10^{-5}$ & $3.31\times 10^{-4}$ & $6.49\times 10^{-5}$ \\
$IV$ & $1.61\times 10^{-7}$ & $2.6\times 10^{-7}$ & $2.07\times 10^{-7}$ & $-8.94\times 10^{-7}$ & $-1.26\times 10^{-6}$ & $-1.2\times 10^{-6}$ \\
\hline
$QI$ & $7.28\times 10^{-8}$ & $-6.1\times 10^{-8}$ & $6.13\times 10^{-8}$ & $1.96\times 10^{-5}$ & $3.34\times 10^{-4}$ & $6.44\times 10^{-5}$ \\
$QQ$ & $0.968$ & $0.976$ & $0.972$ & $5.19\times 10^{-3}$ & $-5.68\times 10^{-3}$ & $-8.66\times 10^{-3}$ \\
$QU$ & $0.968$ & $0.976$ & $0.972$ & $5.18\times 10^{-3}$ & $-5.68\times 10^{-3}$ & $-8.66\times 10^{-3}$ \\
$QV$ & $-1.29\times 10^{-8}$ & $-6.27\times 10^{-8}$ & $-3.63\times 10^{-8}$ & $0.0146$ & $0.0136$ & $0.0125$ \\
\hline
$UI$ & $-1.03\times 10^{-7}$ & $8.19\times 10^{-8}$ & $8.79\times 10^{-8}$ & $2.13\times 10^{-5}$ & $3.32\times 10^{-4}$ & $6.4\times 10^{-5}$ \\
$UQ$ & $0.968$ & $0.976$ & $0.972$ & $5.18\times 10^{-3}$ & $-5.68\times 10^{-3}$ & $-8.66\times 10^{-3}$ \\
$UU$ & $0.968$ & $0.976$ & $0.972$ & $5.19\times 10^{-3}$ & $-5.68\times 10^{-3}$ & $-8.66\times 10^{-3}$ \\
$UV$ & $-9.72\times 10^{-8}$ & $3.84\times 10^{-8}$ & $-4.23\times 10^{-8}$ & $0.0146$ & $0.0136$ & $0.0125$ \\
\hline
$VI$ & $1.77\times 10^{-7}$ & $2.79\times 10^{-7}$ & $1.71\times 10^{-7}$ & $-8.37\times 10^{-7}$ & $-1.54\times 10^{-6}$ & $-7.47\times 10^{-7}$ \\
$VQ$ & $6.71\times 10^{-8}$ & $-1.52\times 10^{-7}$ & $2.55\times 10^{-8}$ & $0.0146$ & $0.0136$ & $0.0125$ \\
$VU$ & $2.02\times 10^{-7}$ & $-1.52\times 10^{-8}$ & $5.24\times 10^{-9}$ & $0.0146$ & $0.0136$ & $0.0125$ \\
$VV$ & $1.95\times 10^{-7}$ & $2.03\times 10^{-7}$ & $2.05\times 10^{-7}$ & $-1.32\times 10^{-6}$ & $2.14\times 10^{-7}$ & $-2.54\times 10^{-6}$ \\
\hline
\end{tabular}

\vspace{2mm}

\begin{tabular}{c|ccc|ccc}
\hline
& \multicolumn{3}{c|}{$c$} & \multicolumn{3}{c}{$d$} \\
$XY$ & Model-A & Model-B & Model-C & Model-A & Model-B & Model-C \\
\hline
$II$ & $4.21\times 10^{-6}$ & $6.79\times 10^{-6}$ & $1.09\times 10^{-6}$ & $-4.5\times 10^{-7}$ & $3.71\times 10^{-7}$ & $1.34\times 10^{-6}$ \\
$IQ$ & $4.26\times 10^{-6}$ & $5.17\times 10^{-5}$ & $3.11\times 10^{-5}$ & $-7.58\times 10^{-6}$ & $-2.76\times 10^{-5}$ & $-2.46\times 10^{-5}$ \\
$IU$ & $4.65\times 10^{-6}$ & $5.36\times 10^{-5}$ & $3.03\times 10^{-5}$ & $-7.4\times 10^{-6}$ & $-2.83\times 10^{-5}$ & $-2.4\times 10^{-5}$ \\
$IV$ & $7.39\times 10^{-6}$ & $6.3\times 10^{-6}$ & $9.41\times 10^{-6}$ & $1.54\times 10^{-6}$ & $-4.52\times 10^{-6}$ & $-2.1\times 10^{-6}$ \\
\hline
$QI$ & $8.37\times 10^{-7}$ & $4.86\times 10^{-5}$ & $3.05\times 10^{-5}$ & $-5.13\times 10^{-6}$ & $-2.54\times 10^{-5}$ & $-2.38\times 10^{-5}$ \\
$QQ$ & $4.69\times 10^{-4}$ & $-1.66\times 10^{-4}$ & $-6.11\times 10^{-5}$ & $-1.27\times 10^{-4}$ & $1.71\times 10^{-4}$ & $3.54\times 10^{-4}$ \\
$QU$ & $3.7\times 10^{-4}$ & $-2.59\times 10^{-4}$ & $-1.39\times 10^{-4}$ & $-1.32\times 10^{-4}$ & $1.73\times 10^{-4}$ & $3.54\times 10^{-4}$ \\
$QV$ & $7.21\times 10^{-5}$ & $-2.22\times 10^{-4}$ & $-5.51\times 10^{-5}$ & $1.44\times 10^{-6}$ & $2.61\times 10^{-6}$ & $5.25\times 10^{-6}$ \\
\hline
$UI$ & $-3.53\times 10^{-6}$ & $5.27\times 10^{-5}$ & $3.14\times 10^{-5}$ & $-2.08\times 10^{-6}$ & $-2.81\times 10^{-5}$ & $-2.43\times 10^{-5}$ \\
$UQ$ & $3.7\times 10^{-4}$ & $-2.59\times 10^{-4}$ & $-1.39\times 10^{-4}$ & $-1.33\times 10^{-4}$ & $1.73\times 10^{-4}$ & $3.53\times 10^{-4}$ \\
$UU$ & $2.54\times 10^{-4}$ & $-3.49\times 10^{-4}$ & $-2.18\times 10^{-4}$ & $-1.28\times 10^{-4}$ & $1.75\times 10^{-4}$ & $3.54\times 10^{-4}$ \\
$UV$ & $7.5\times 10^{-5}$ & $-2.16\times 10^{-4}$ & $-5.64\times 10^{-5}$ & $-2.15\times 10^{-6}$ & $-2.23\times 10^{-6}$ & $5.17\times 10^{-6}$ \\
\hline
$VI$ & $7.28\times 10^{-6}$ & $7\times 10^{-6}$ & $8.32\times 10^{-6}$ & $1.56\times 10^{-6}$ & $-4.94\times 10^{-6}$ & $-1.42\times 10^{-6}$ \\
$VQ$ & $7.55\times 10^{-5}$ & $-2.21\times 10^{-4}$ & $-5.5\times 10^{-5}$ & $-2.65\times 10^{-7}$ & $2.23\times 10^{-6}$ & $5.65\times 10^{-6}$ \\
$VU$ & $8.04\times 10^{-5}$ & $-2.17\times 10^{-4}$ & $-5.99\times 10^{-5}$ & $-5.57\times 10^{-6}$ & $-1.25\times 10^{-6}$ & $8.13\times 10^{-6}$ \\
$VV$ & $2.16\times 10^{-4}$ & $1.9\times 10^{-4}$ & $1.64\times 10^{-4}$ & $2.49\times 10^{-6}$ & $-4.6\times 10^{-6}$ & $-6.65\times 10^{-6}$ \\
\hline
\end{tabular}
\end{table*}


\subsection{Dependence on frequency and band average}
\label{subsec:bandave}
\begin{figure}
    \centering
    \includegraphics[width=1.0\linewidth]{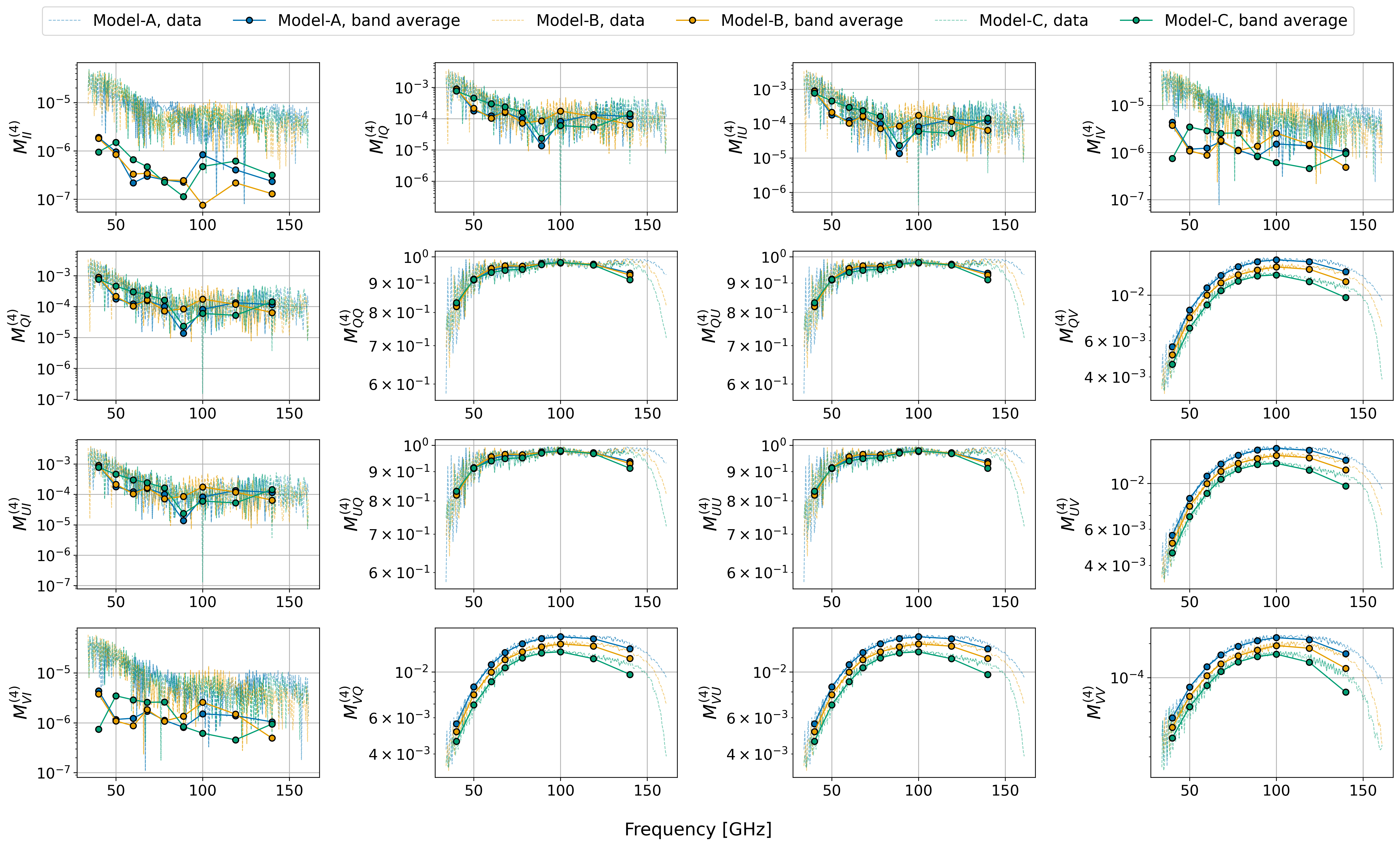}
    \caption{The amplitude $M_{XY}^{(4)}$ as a function of frequency $\nu$ with an incident angle of 10 degrees. The dashed line shows the RCWA calculation, and dot data shows their band-average for each LFT band. The blue, orange and green colored data show the results for Model- A, B and C, respectively.}
    \label{fig:bandave_amplitude}
\end{figure}

\begin{figure}
    \centering
    \includegraphics[width=1.0\linewidth]{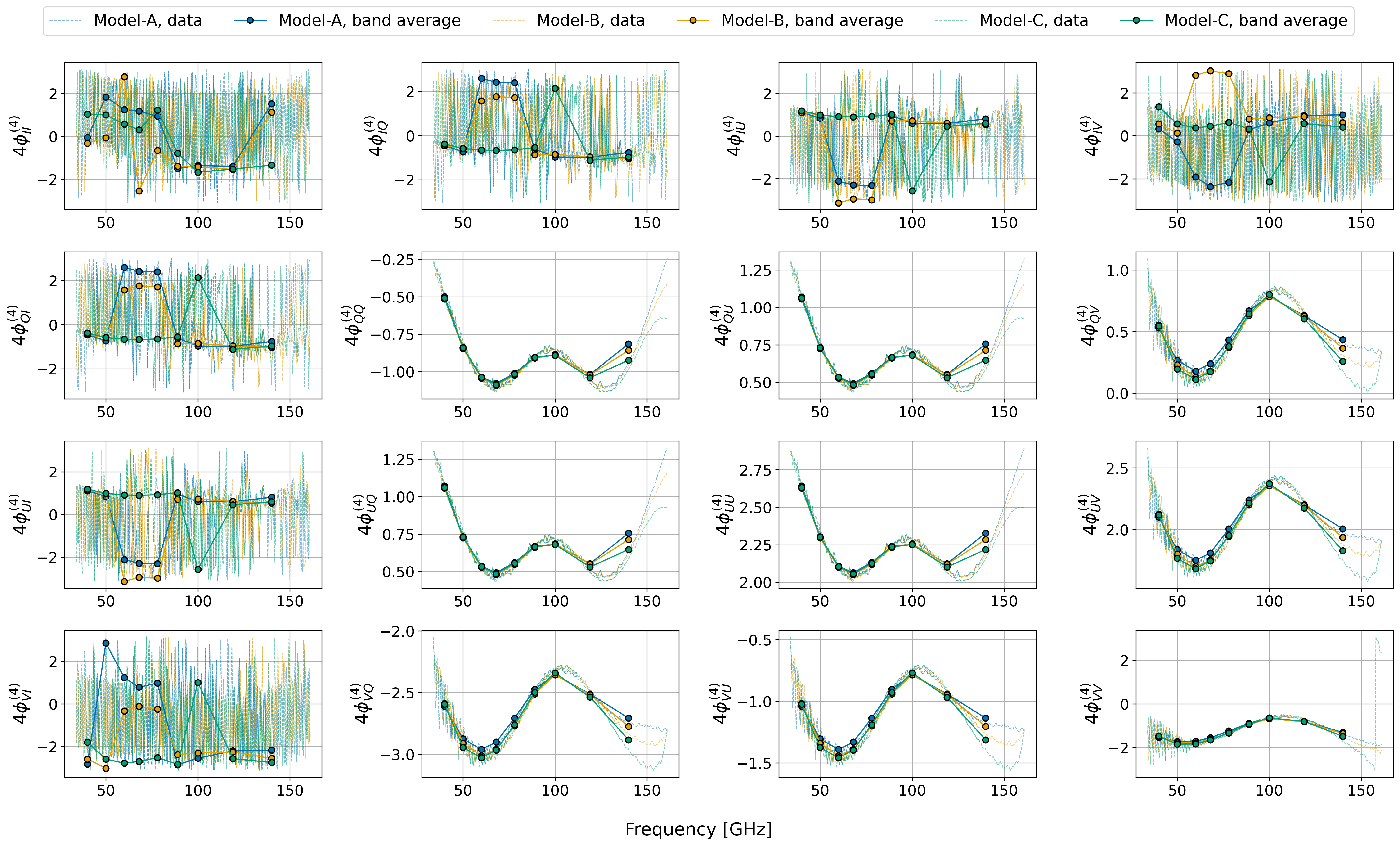}
    \caption{The phase $\phi_{XY}^{(4)}$ as a function of frequency $\nu$ with an incident angle of 10 degrees. The axes, colors and markers indicate the same with Figure \ref{fig:bandave_amplitude}.}
    \label{fig:bandave_phase}
\end{figure}

Figure~\ref{fig:bandave_amplitude} shows the amplitudes of
$M_{XY}^{(4)}(\Theta = 10^{\circ})$ as a function of frequency $\nu$,
together with their band-averaged values.
The corresponding phases
$\phi_{XY}^{(4)}(\Theta = 10^{\circ})$ are shown in
Figure~\ref{fig:bandave_phase}.
The band averages of both the amplitude and
phase are calculated with Equation~\ref{eq:bandavg_complex_mueller}.

A thicker plate and a higher refractive index produce faster oscillations in the frequency domain.
This behavior is visible in the amplitude spectra in Figure ~\ref{fig:bandave_amplitude}.
At the same time, Figure~\ref{fig:bandave_phase} shows that the phases of $M_{XI/IX}^{(4)}$ $(X=Q,U,V)$ vary rapidly within each band.
This can lead to the cancellation of $M_{XI/IX}^{(4)}$ when they are averaged over frequency. 
We can see that the cancellation happens in each panel for $M_{XI/IX}^{(4)}$ in Figure~\ref{fig:bandave_amplitude}.
The resulting band-averaged amplitudes and phases are also consistent with the
relations expected from the theoretical symmetry of the fourth-harmonic terms (see appendix A of Patanchon et al~\cite{Patanchon_2024}.):

\begin{align}
\label{eq:relation_epsilon_phi}
        \begin{cases}
            \varepsilon_{IQ}^{(4)} = \varepsilon_{IU}^{(4)}, \ \ \ \ \ \  4\phi_{IQ}^{(4)} = 4\phi_{IU}^{(4)} + \pi/2\\
            \varepsilon_{QI}^{(4)} = \varepsilon_{UI}^{(4)}, \ \ \ \ \ \  4\phi_{QI}^{(4)} = 4\phi_{UI}^{(4)} + \pi/2\\
            \varepsilon_{QV}^{(4)} = \varepsilon_{UV}^{(4)}, \ \ \ \ \ \  4\phi_{QV}^{(4)} = 4\phi_{UV}^{(4)} + \pi/2\\
            \varepsilon_{VQ}^{(4)} = \varepsilon_{VU}^{(4)}, \ \ \ \ \ \  4\phi_{VQ}^{(4)} = 4\phi_{VU}^{(4)} + \pi/2\\
        \end{cases}
\end{align}

\begin{figure}
    \centering
    \includegraphics[width=1.0\linewidth]{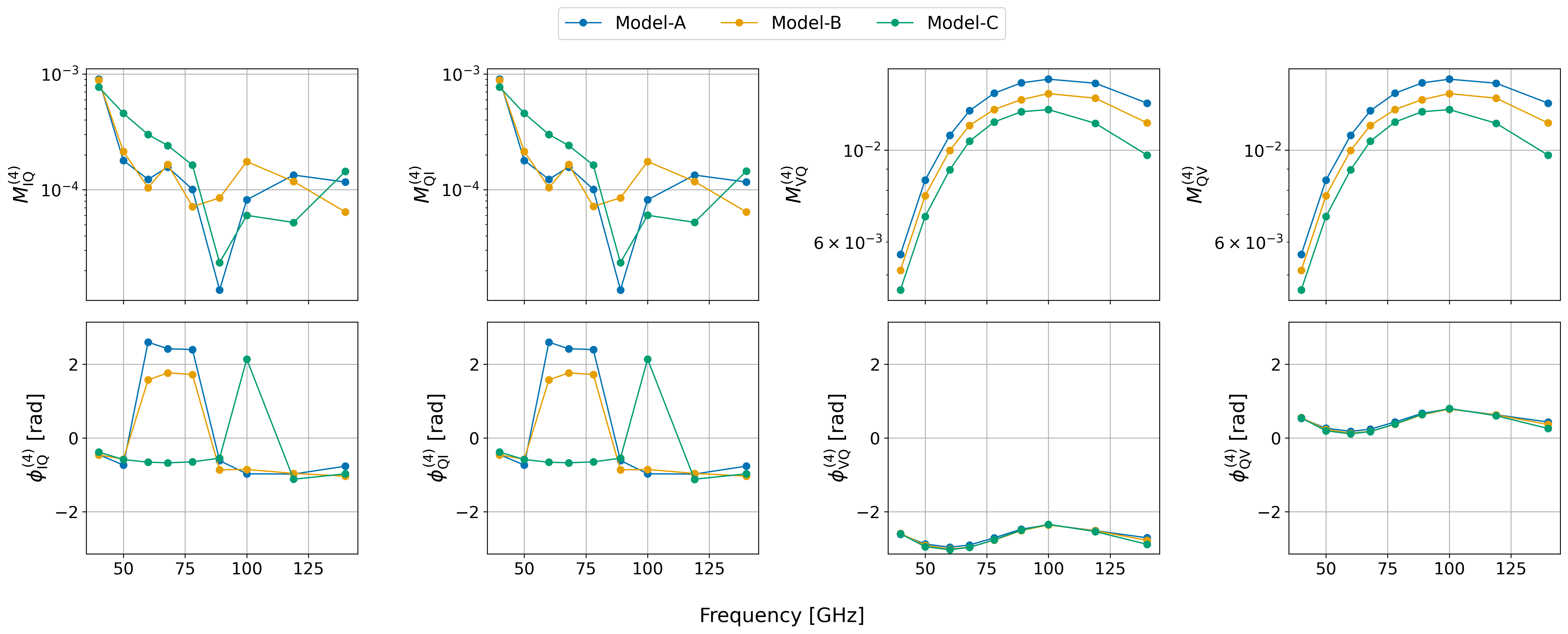}
    \caption{The band-averaged amplitudes and phases at the fourth harmonics. The axes, colors and markers indicate the same with Figure \ref{fig:bandave_amplitude}.}
    \label{fig:IQQIVQQV}
\end{figure}

The band-averaged amplitudes and phases of Mueller coefficient $4f$ harmonics
 denoted by $\varepsilon_{XY}^{(4)}$ and $\phi_{XY}^{(4)}$, are complied in Figure~\ref{fig:IQQIVQQV}. 
The difference of the amplitude $\varepsilon_{IQ}^{(4)} - \varepsilon_{IU}^{(4)}$ and $\varepsilon_{QI}^{(4)} - \varepsilon_{UI}^{(4)}$ is less than $2.0\times 10^{-7}$, while $\varepsilon_{VQ}^{(4)} - \varepsilon_{VU}^{(4)}$ and $\varepsilon_{QV}^{(4)} - \varepsilon_{UV}^{(4)}$ is less than $2.0\times 10^{-6}$. 
These small differences indicate that the numerical results are consistent with the expected relations in Equation~\ref{eq:relation_epsilon_phi}.
The order of these differences is compatible with numerical precision within three SWS models.
The difference of the phase shown in Equation ~\ref{eq:relation_epsilon_phi} is less than $1.0\times 10^{-3}$ rad. 
The instrumental polarization $M_{IX}^{(4)}(\nu)$ and $M_{XI}^{(4)}(\nu)$ ($X = Q,U$) is the order of $10^{-5}$ at higher frequency, while reaches $10^{-3}$ at lower frequency bands for all SWS models.

Since the simulations have been run at every 0.5~GHz, we evaluate the dependence with the band coverage by shifting the integration range by up to 1 GHz. The differences remain at approximately $\mathcal{O}(10^{-5})$ or below for most frequency bands and models, indicating that such shifts in the band-pass range have a negligible impact on the band-averaged instrumental polarization.

\begin{figure}
    \centering
    \includegraphics[width=1.0\linewidth]{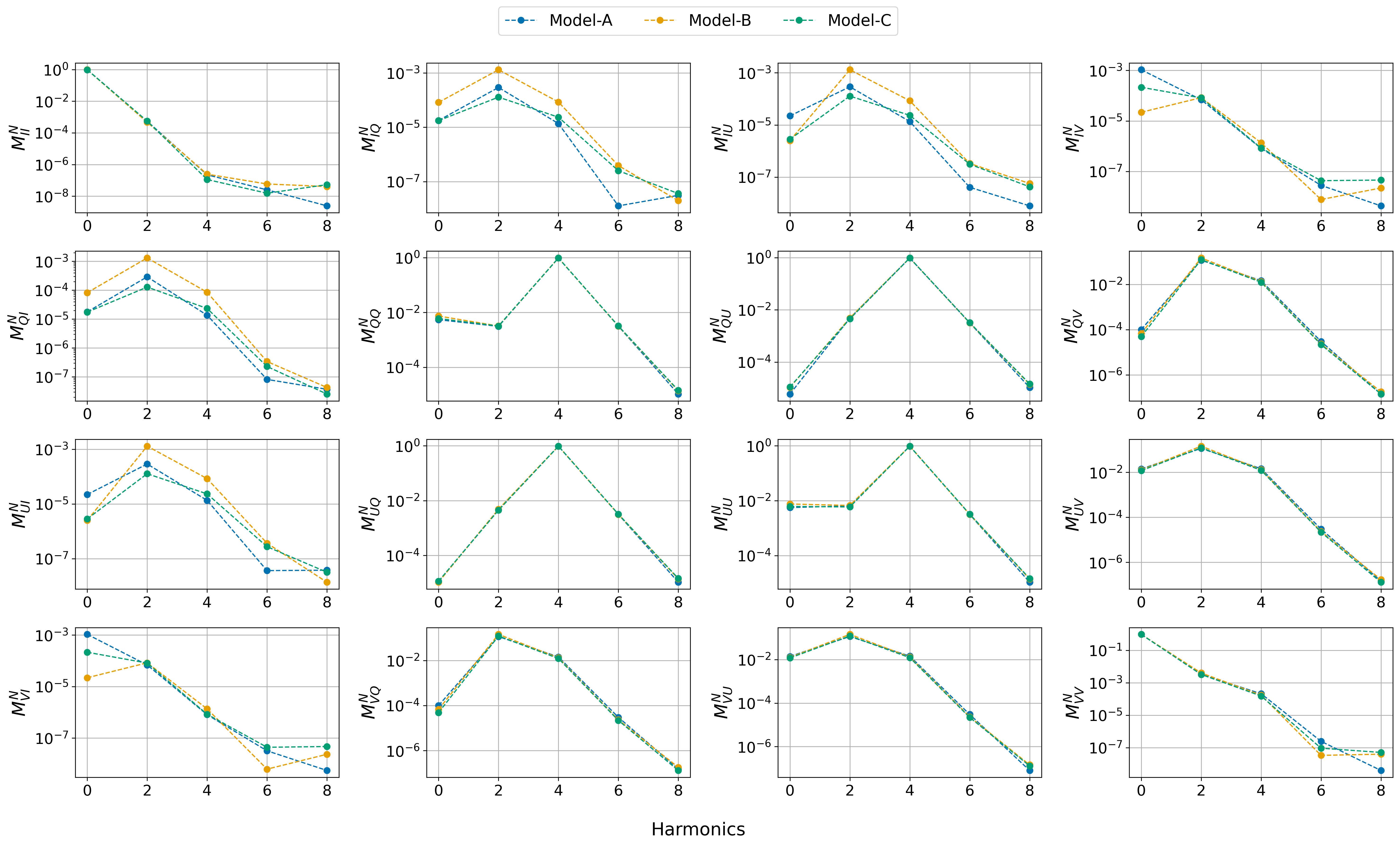}
    \caption{The amplitudes of the Mueller matrix averaged over frequency centered at 89 GHz for an incident angle of 10 degrees. blue, orange, green lines show Model- A, B and C, respectively.}
    \label{fig:harmonics_89GHz}
\end{figure}

Although we focus on the fourth harmonics, we also make plots of the magnitudes of the order harmonics up to the eighth e.g. at 89~GHz in Figure~\ref{fig:harmonics_89GHz}. 
The conversion between linear and circular polarization are relatively large and at the order of $10^{-2}$, which is because the oblique incidence gives rise to the change in the optical path length. 
The other harmonics than the fourth one will be needed if a detector is not a linear system. We include only the $0f$, $2f$, $4f$, $6f$ and $8f$ components because higher-order harmonics are negligibly small. We will discuss the detector non-linearity in Section \ref{subsec:non-linearity}.

\subsection{Effect on linear polarization}
\label{subsec:conversion_linear_pol}

\begin{figure}
    \centering
    \includegraphics[width=1.0\linewidth]{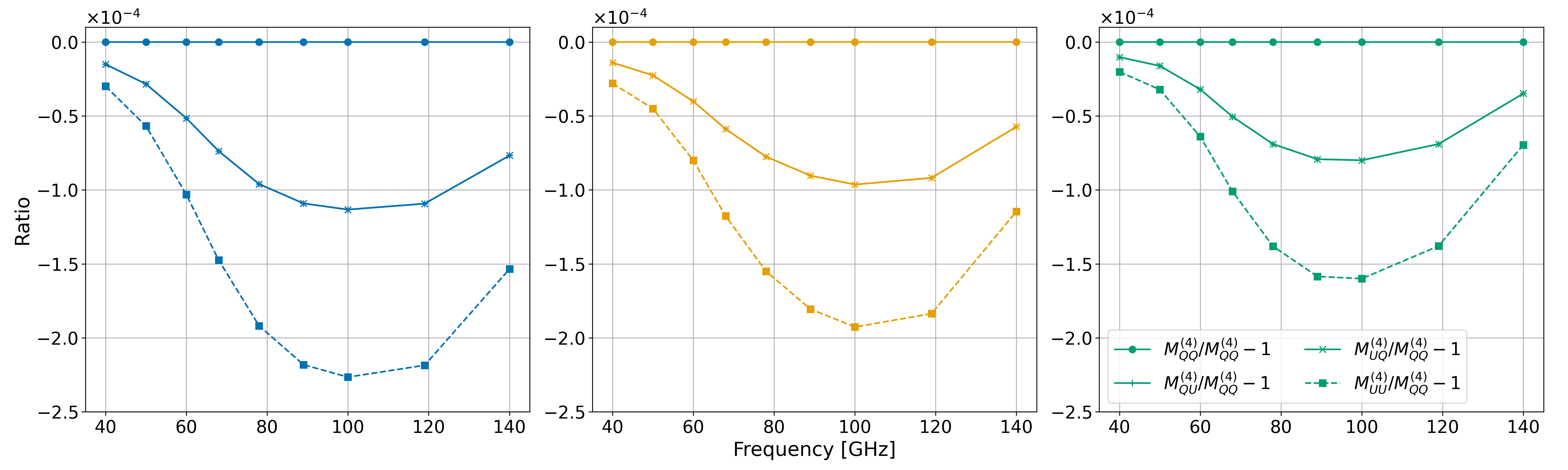}
    \includegraphics[width=1.0\linewidth]{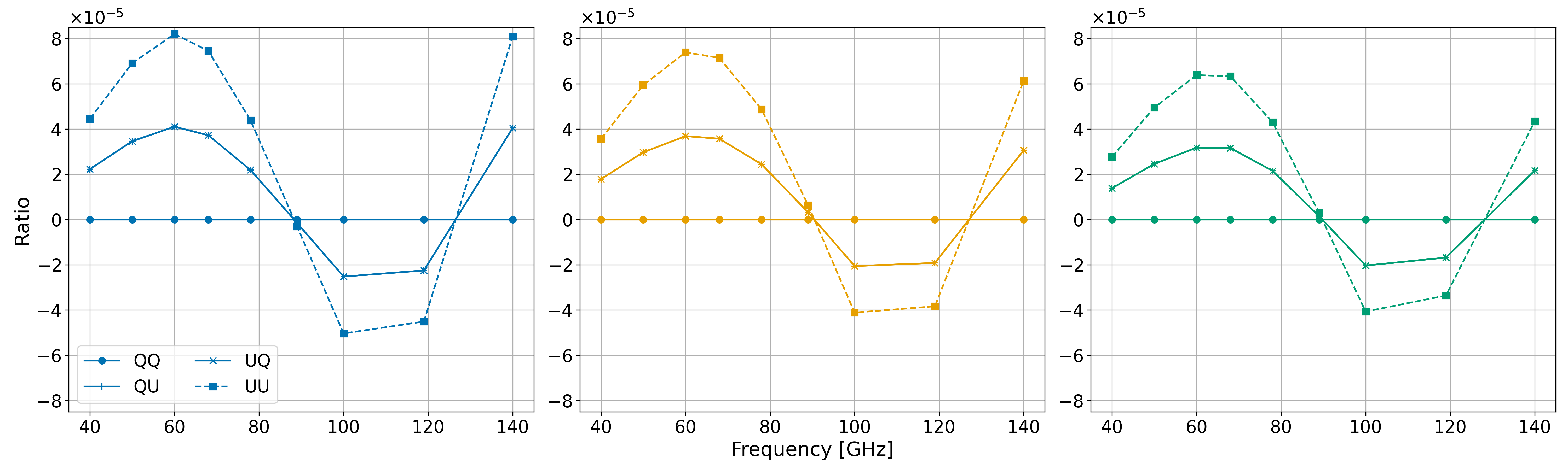}
    \caption{The band-averaged amplitudes (top) and phases (bottom) for three models(left: Model-A, center: Model-B, right: Model-C.)}
    \label{fig:linear_pols}
\end{figure}

We find unintended conversion between Stokes $Q$ and $U$ unless the following relations hold precisely: 
$\varepsilon^{(4)}_{QQ} = \varepsilon^{(4)}_{QU} = \varepsilon^{(4)}_{UQ} = \varepsilon^{(4)}_{UU}$, and $4\phi^{(4)}_{QQ} = 4\phi^{(4)}_{QU} - \pi/2 = 4\phi^{(4)}_{UQ} - \pi/2 = 4\phi^{(4)}_{UU} - \pi$. The simulation demonstrates that these relations break slightly, as shown in Figure~\ref{fig:linear_pols}. The differences are defined as

\begin{align}
    \Delta_{XY} &= \frac{\varepsilon^{(4)}_{XY}}{\varepsilon^{(4)}_{QQ}} - 1 \nonumber\\
    \delta_{XY}
                &= 4\varphi_{XY}^{(4)} - 4\varphi_{QQ}^{(4)}
                - \begin{cases}
                \pi   & \left( XY \in \{UU\} \right) \\
                \pi/2 & \left( XY \in \{QU, UQ\} \right)
                \end{cases}
\end{align}

\begin{figure}
    \centering
    \includegraphics[width=0.7\linewidth]{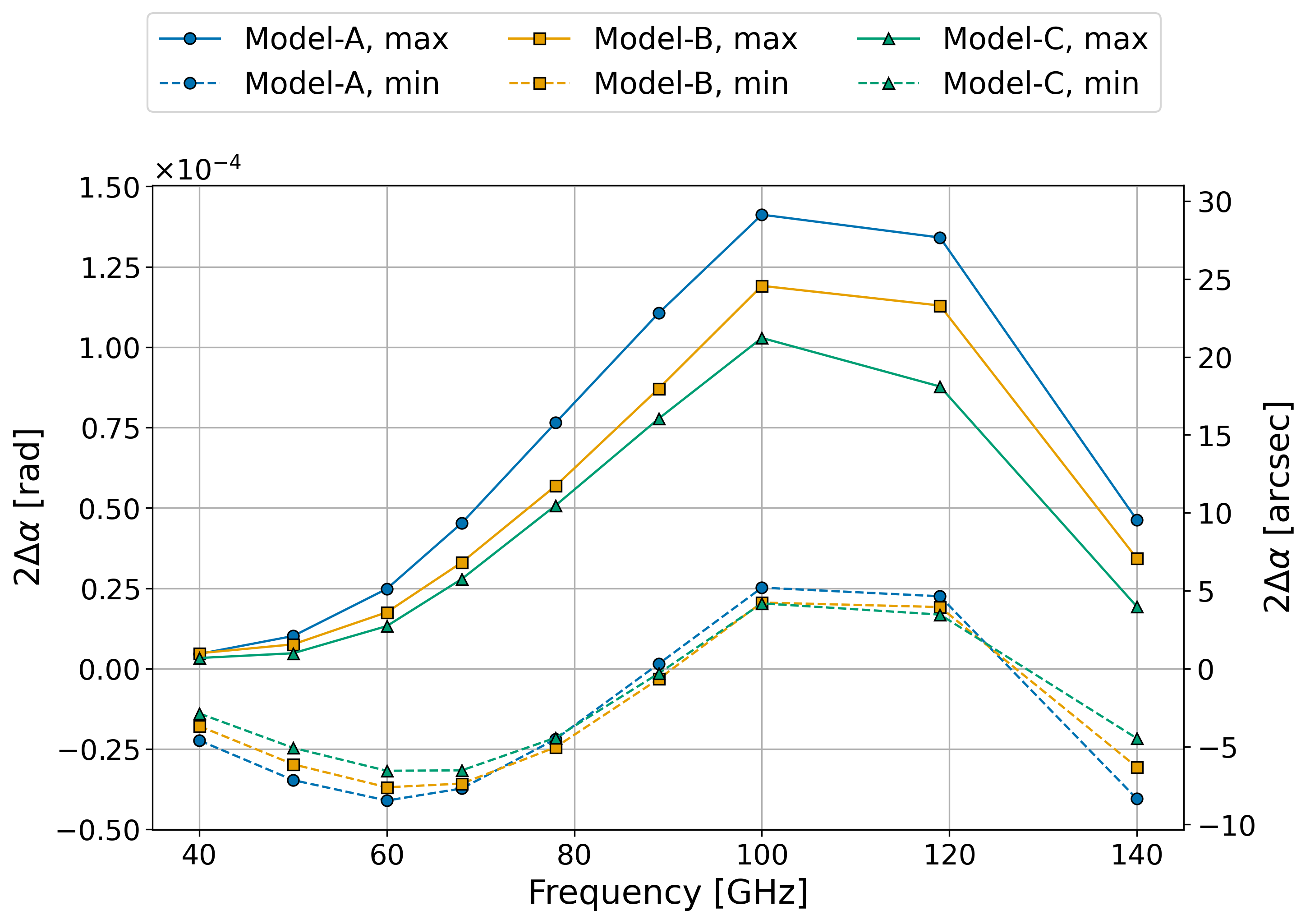}
    \caption{Maximum and minimum additional angle for three models, with the unit of radian and arcseconds.}
    \label{fig:2deltaalpha}
\end{figure}

These differences add shift in the polarization angle as well as a polarization efficiency coefficient. The additional angle depends on the incident polarization angle. 
The additional angles are summarized in Figure~\ref{fig:2deltaalpha}.
The maximum additional angle is $1.4\times10^{-4}$, and minimum additional angle is $-4.0\times10^{-5}$, corresponding to a maximum absolute rotation of approximately 30 arcsec.
Appendix B describes how to estimate the additional angle rotation. To detect the primordial B-mode signals corresponding to a tensor-to-scalar ratio of $r=0.001$, recent studies have shown that the required polarization-angle accuracy depends on the frequency band and the correlation of angle errors among detectors, with the most stringent requirement being approximately one arcmin.~\cite{Vielva_2022}. The maximum additional rotation found here, approximately 30 arcsec, is below this level.

\subsection{Transmittance and polarization efficiency}
\label{subsec:trans_modeff}

\begin{figure}
    \centering
    \includegraphics[width=1.0\linewidth]{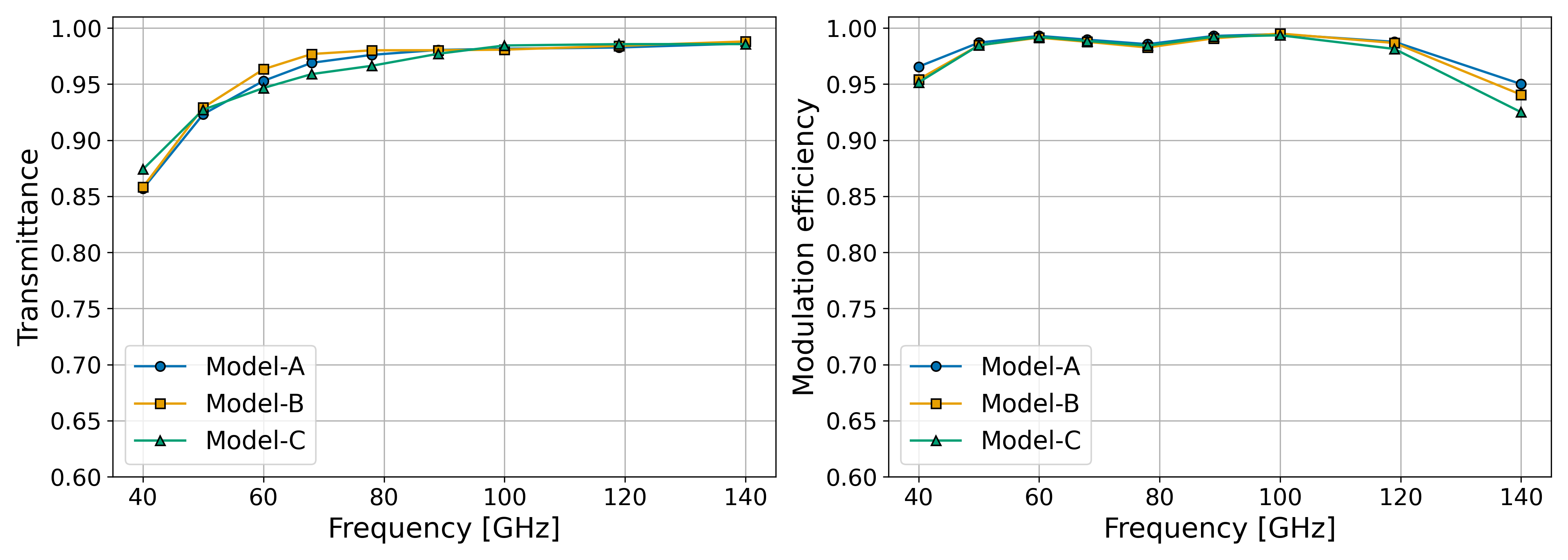}
    \caption{Band-averaged transmittance (left) and polarization efficiency (right) for three models.}
    \label{fig:trans_modeff}
\end{figure}

In the previous sections, we focused on the polarization conversion induced by the HWP mainly.
However, there are other parameters that characterize the performance of the HWP.
These are transmittance and polarization efficiency, as described in Sec.~\ref{subsec:sim_outputs}. 
Figure~\ref{fig:trans_modeff} shows the transmittance and polarization efficiency for three models. 
These parameters do not much depend on the SWS shape, and the transmittance is greater than 95\% except lower two frequency bands, and polarization efficiency is greater than 95\% except the highest frequency band. 5\% for the transmittance includes the multiple reflections between HWP layers, thus SWS on the first and last plates work well at higher frequency bands.
In order to improve the lower transmittance at the lower frequency bands, we need to improve the shape of the structures or to get deeper structures.

\subsection{Realistic SWS}
\label{subsec:realistic_case}

\begin{figure}
    \centering
    \includegraphics[width=0.4\linewidth]{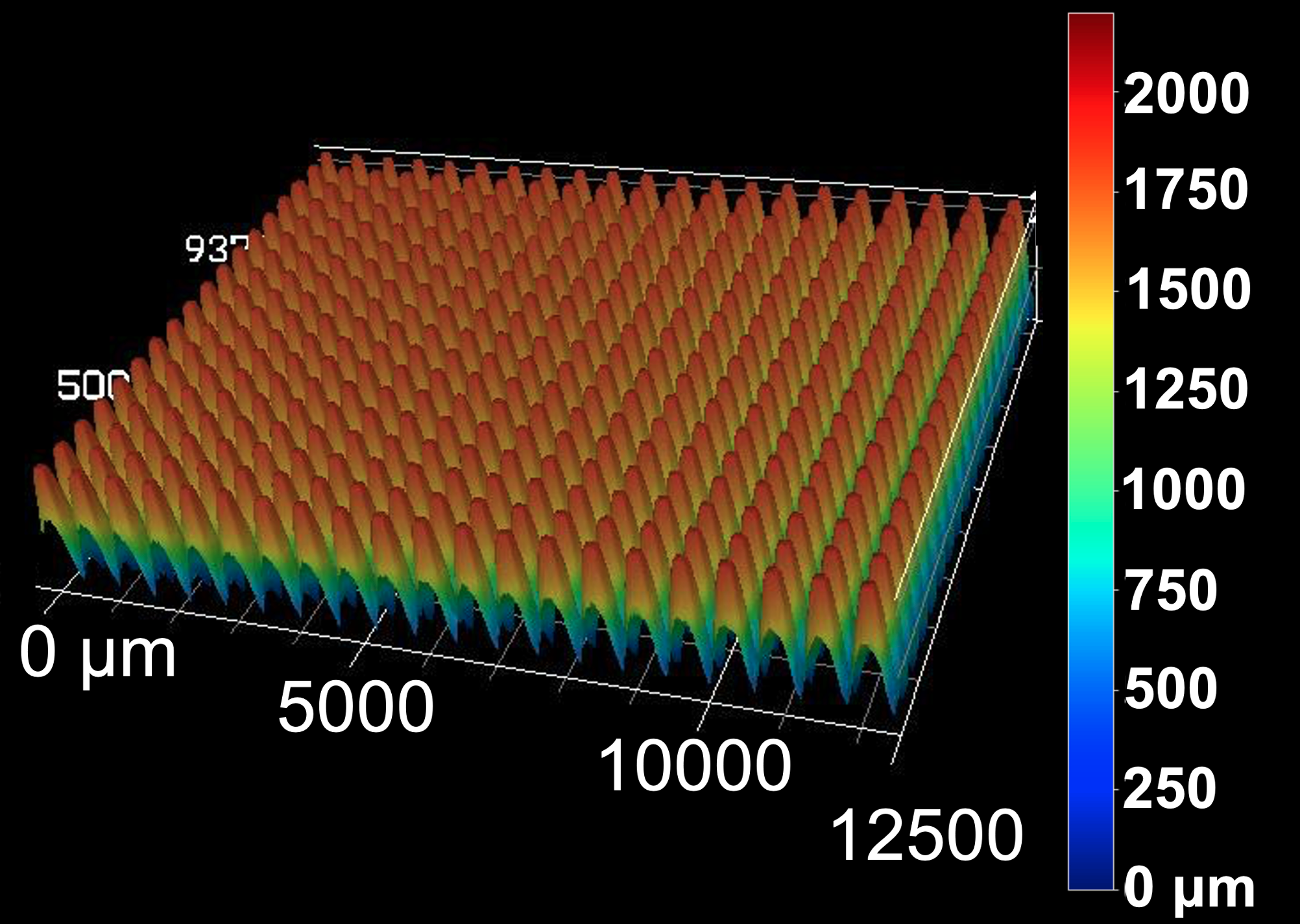}
    \includegraphics[width=0.5\linewidth]{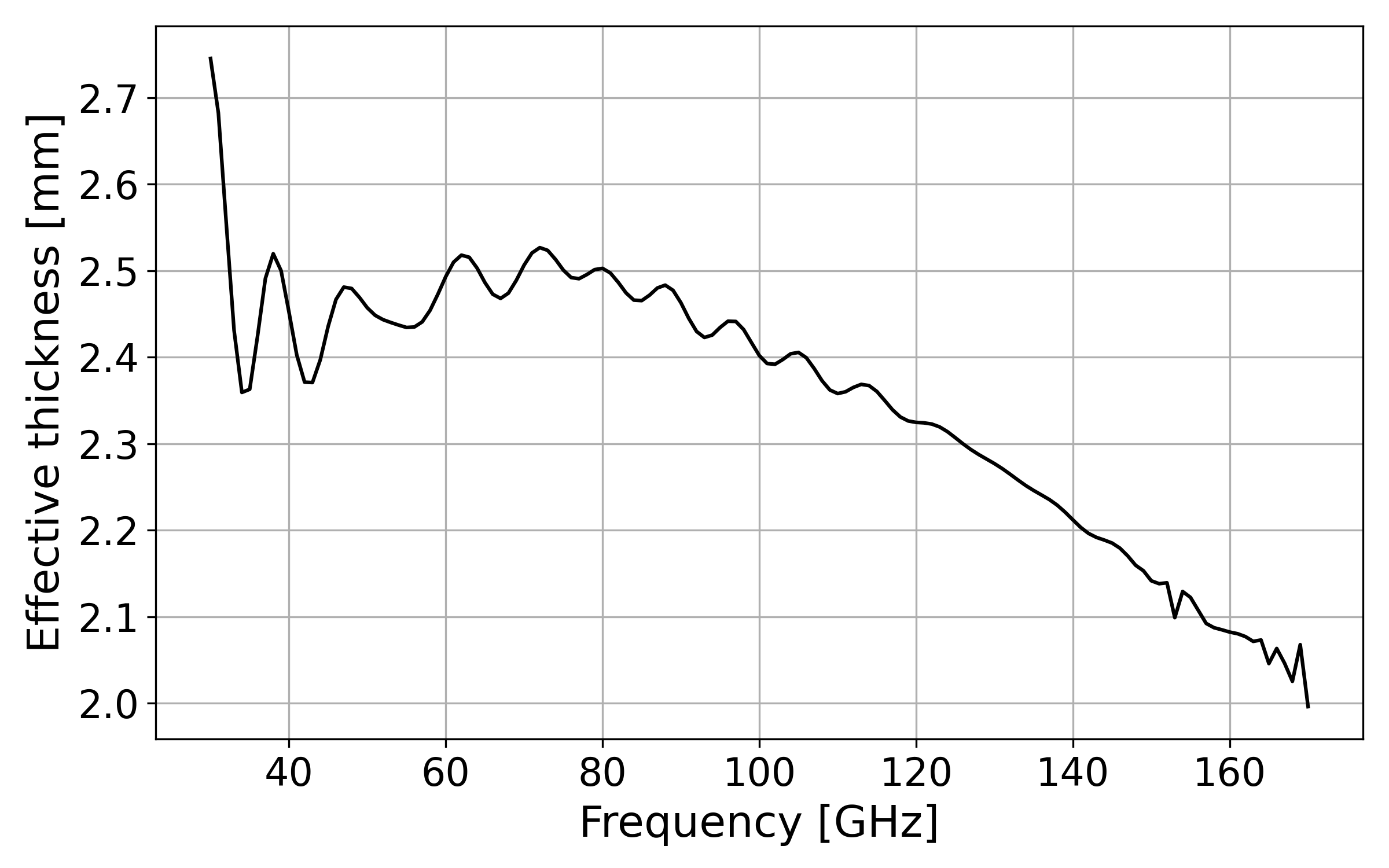}
    \caption{Image of the fabricated structures taken by con-focal microscope (left) from Takaku et al (2023), and effective thickness (right). }
    \label{fig:carbide18_image_teff}
\end{figure}

\begin{figure}
    \centering
    \includegraphics[width=1.0\linewidth]{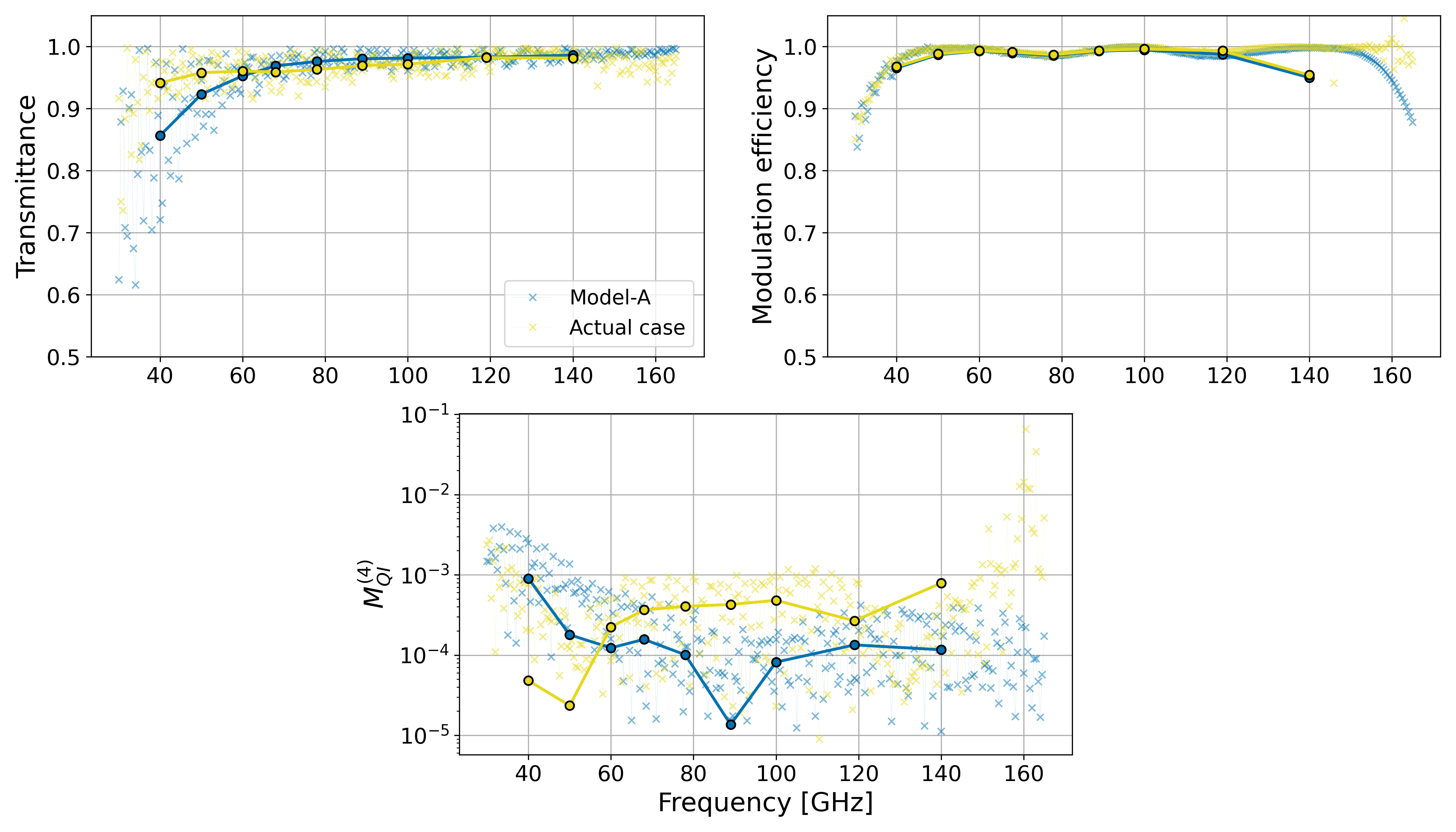}
    \caption{Transmittance $M_{II}^{(0)}$ (top left), polarization efficiency $M_{QQ}^{(4)} / M_{II}^{(0)}$ (top right), and $M_{XY}^{(4)}$ (bottom) of the realistic case. As a reference we also show these parameters of the Model-A.}
    \label{fig:carbide18_spectrum}
\end{figure}

Takaku et al.~\cite{Takaku_SPIE2022} reports on the measurement performance of sapphire based HWP with SWS fabricated with ultra-short pulse laser ablation. 
The structures are fabricated on each of the two outer 200-mm-diameter sapphire plates, and the geometry type follows the Model-A.
As the estimation of the HWP coefficient for a realistic structure, we use a structure on Plate 1, with the total height of 2.12 mm and pitch in $x$ and $y$ is 0.6 mm, 0.54 mm, respectively.
Fig.~\ref{fig:carbide18_image_teff} shows the con-focal image and its effective thickness. 
We tuned the bulk thickness of the two outer plates as 2.528 mm when we calculate the Mueller matrix as a function of frequency for an incident angle of 10 degrees.
Fig. ~\ref{fig:carbide18_spectrum} shows the transmittance $M_{II}^{(0)}$, polarization efficiency $M_{QQ}^{(4)} / M_{II}^{(0)}$, and $M_{XY}^{(4)}$ quantifying instrumental polarization.
The polarization efficiency is similar to Model-A. 
The transmittance and IP appear to be anti-correlated, in higher transmittance is associated to lower IP. 
In the realistic case, the structure is deeper than for the model, which results in higher transmittance at low frequencies. 
However, the transmittance at the middle frequency bands decreases compared with the model. 
The top width of the realistic case is approximately 50 $\mathrm{\mu}$m larger than the model, which leads the surface reflection.

\section{Discussion}
\label{sec:discussion}

\subsection{Color effects and foreground}
\label{subsec:foregrounds}

We calculated the averaged $M^{(4)}_{XY}$ over frequency bands with the CMB temperature anisotropy and estimated the contamination of the $4\rho$ signal in Section~\ref{subsec:bandave}. 
The sky signal contains not only the CMB but also galactic foreground emissions, 
which can significantly affect CMB polarization observations. 
Galactic dust and synchrotron emissions are intrinsically polarized. 
However, in this analysis, we focus on the spurious polarization generated by the instrumental response to their unpolarized intensity components. 
Thus, we model the dust and synchrotron intensity spectra as a modified blackbody and a power law, respectively, using representative spectral-index values.

We calculated the band average for both dust and synchrotron emissions with

\begin{equation}
\varepsilon_{\mathrm{dust},XY}^{(2i)}
=
\left|
\frac{
\displaystyle \int_{\nu_1}^{\nu_2}
d\nu\,
M_{XY}^{(2i)}(\nu)\,
e^{\,i\phi_{XY}^{(2i)}(\nu)}\,
B(\nu,T_d)
\left(\frac{\nu}{\nu_0}\right)^{\beta_d}
}{
\displaystyle \int_{\nu_1}^{\nu_2}
d\nu\,
B(\nu,T_d)
\left(\frac{\nu}{\nu_0}\right)^{\beta_d}
}
\right|
\end{equation}

\begin{equation}
\varepsilon_{\mathrm{sync},XY}^{(2i)}
=
\left|
\frac{
\displaystyle \int_{\nu_1}^{\nu_2}
d\nu\,
M_{XY}^{(2i)}(\nu)\,
e^{\,i\phi_{XY}^{(2i)}(\nu)}\,
\left(\frac{\nu}{\nu_0}\right)^{\beta_s}
}{
\displaystyle \int_{\nu_1}^{\nu_2}
d\nu\,
\left(\frac{\nu}{\nu_0}\right)^{\beta_s}
}
\right|
\end{equation}

instead of Eq.~\ref{eq:bandavg_complex_mueller}. 
We adopted $T_d = 19.6\,\mathrm{K}$ and varied $\beta_d$ and $\beta_s$ to investigate the dependence of the band-averaged response on the foreground spectral shapes. 
We also calculated the band-averaged response using representative values for the Planck dust and synchrotron intensity models, namely, $\beta_d = 1.48$ and $\beta_s = -3.1$, respectively~\cite{Planck2018XI,Planck2015X}.

\begin{figure}
    \centering
    \includegraphics[width=1.0\linewidth]{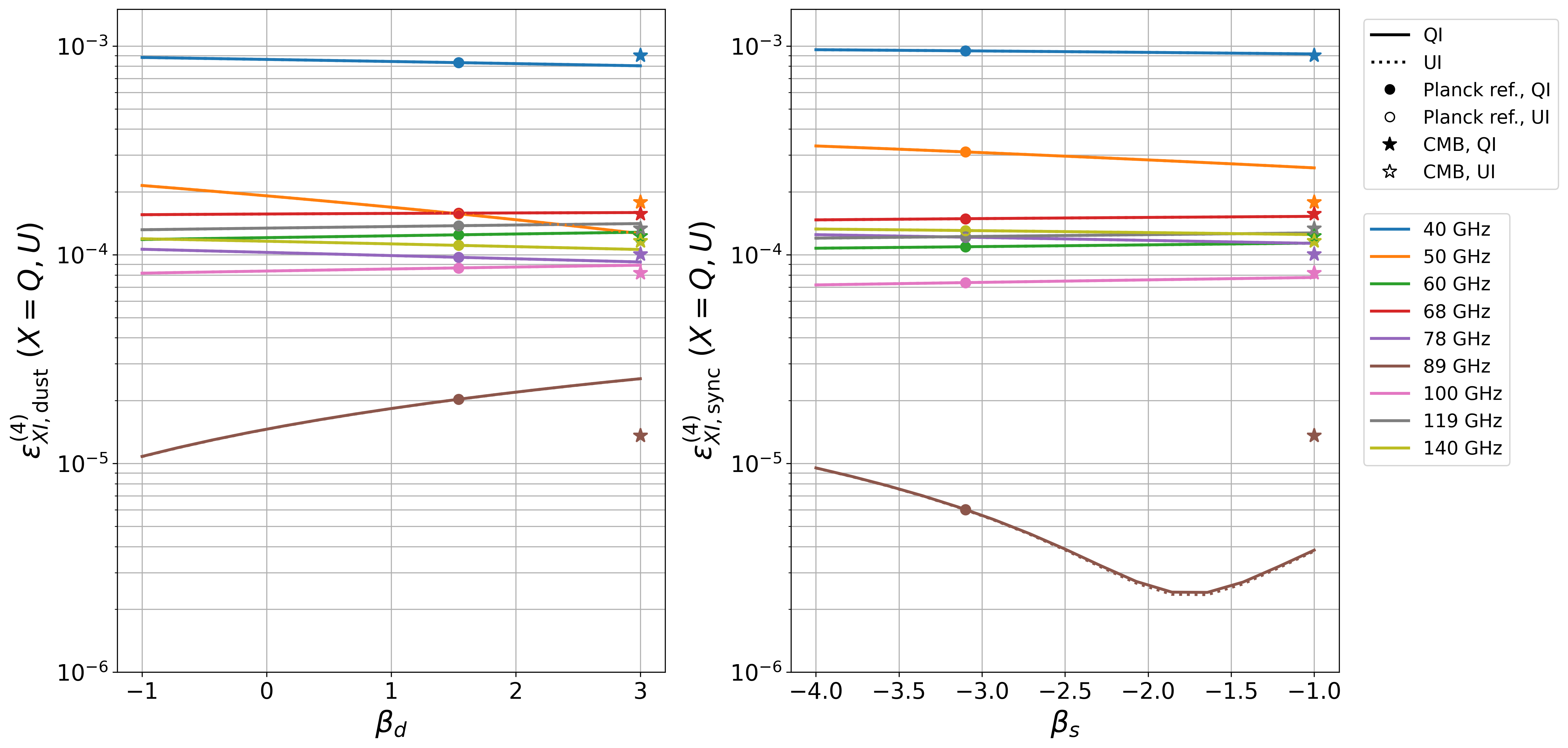}
    \caption{Dependence of the band-averaged $\varepsilon_{QI}^{(4)}$ (solid lines) and $\varepsilon_{UI}^{(4)}$ (dotted lines) on the spectral indices of dust (left) and synchrotron (right) for Model-A. Colors indicate the band center frequencies. The dust temperature is assumed to be 19.6 K.}
    \label{fig:foreground}
\end{figure}

Figure~\ref{fig:foreground} shows the band-averaged values of $\varepsilon_{QI}^{(4)}$ and $\varepsilon_{UI}^{(4)}$. The QI and UI curves nearly overlap. 
The results obtained using the representative Planck values for the dust and synchrotron spectral indices, together with the result for the CMB temperature anisotropy, are shown for reference. 
To assess the sensitivity to spatial variations of the foreground spectral shapes, we vary the spectral indices by $\Delta\beta_d = \pm 0.10$ and $\Delta\beta_s = \pm 0.07$, motivated by representative variations reported in Planck and beyond-Planck analyses~\cite{Planck2018XI,Svalheim2023}. 
The resulting changes in the band-averaged efficiency are less than 1\% for most frequency bands, reaching about 1.4\% and 1.8\% for dust at 50 and 89~GHz, respectively, and about 4.4\% for synchrotron at 89~GHz. 
The relatively large fractional change at 89~GHz for synchrotron is due to the very small absolute value of $\varepsilon_{QI}^{(4)}$ and $\varepsilon_{UI}^{(4)}$.

\subsection{Detector non-linearity}
\label{subsec:non-linearity}
Micheli et al.~\cite{Micheli2026} showed that the coupling between HWP differential emission and TES non-linearity can significantly contaminate CMB polarization reconstruction. Although our study focuses on differential transmission, a similar coupling with detector non-linearity may need to be considered.

Including detector non-linearity, the data model can be written as
\begin{align}
d = S + c_{\mathrm{NL}} S^2 + O(S^3),
\end{align}
where $S$ is the incident power to the detector.
Detector non-linearity can generate a spurious signal at $4f_{\mathrm{rot}}$
through mixing of other harmonic components.
In this subsection, we consider the contributions from the constant,
second, fourth, and sixth harmonics.

We consider the Mueller matrix elements that have the largest values
at each harmonic.
The incident power is then decomposed into
\begin{align}
2S^{(0)} &= \sum \mathcal{M}_{XX}^{(0)} X_{\mathrm{sky}},
&& (X = I, V), \\
2S^{(2)} &= \sum \mathcal{M}_{XY}^{(2)} Y_{\mathrm{sky}},
&& \left( XY \in \{QV, UV, VQ, VU\} \right), \\
2S^{(4)} &= \sum \mathcal{M}_{XY}^{(4)} Y_{\mathrm{sky}},
&& \left( XY \in \{QQ, QU, UQ, UU\} \right), \\
2S^{(6)} &= \sum \mathcal{M}_{XY}^{(6)} Y_{\mathrm{sky}},
&& \left( XY \in \{QQ, QU, UQ, UU\} \right).
\end{align}

We introduce the notation $d^{(4)}$ for the fourth-harmonic component
of the detector output.
To leading order in the quadratic non-linearity, it can be written as
\begin{align}
d^{(4)} \sim S^{(4)} + c_{\mathrm{NL}}
\left[
S^{(0)} S^{(4)}
+ \left( S^{(2)} \right)^2
+ S^{(2)} S^{(6)}
\right].
\end{align}

Since $S^{(0)}$ is a constant component, the term
$c_{\mathrm{NL}}S^{(0)}S^{(4)}$ changes only the amplitude of the
original $S^{(4)}$ component without changing its phase.
Therefore, this contribution can in principle be absorbed into an
amplitude calibration.
On the other hand, the other two terms can modify both the amplitude
and phase of the fourth-harmonic signal.

Figure~\ref{fig:harmonics_89GHz} shows the orders of magnitude of the
band-averaged Mueller matrix components as a function of harmonic
order.
Takakura~et~al.~\cite{Takakura_2017} estimated the responsivity
non-linearity of TES bolometers to be approximately
$-0.17\%\,\mathrm{K}^{-1}$ or $-0.42\%\,\mathrm{K}^{-1}$ for POLARBEAR.
We identify this responsivity non-linearity with $c_{\mathrm{NL}}$
in our quadratic model and conservatively adopt
$|c_{\mathrm{NL}}|= 4.2\times10^{-3}\,\mathrm{K}^{-1}$.
Here, we consider only the responsivity non-linearity and neglect
the non-linearity associated with the detector time constant.
From Fig.~\ref{fig:harmonics_89GHz}, representative values of the
harmonic components are
\begin{align}
S^{(2)} &\sim 0.1 P_{\mathrm{sky}}, \\
S^{(4)} &\sim P_{\mathrm{sky}}, \\
S^{(6)} &\sim 0.003 P_{\mathrm{sky}}.
\end{align}
Using these values, the harmonic-mixing terms are estimated as
\begin{align}
|c_{\mathrm{NL}}| \left(S^{(2)}\right)^2
&\sim 4\times10^{-11}~
\mu\mathrm{K}^{-1} \cdot P_{\mathrm{sky}}^{\,2}, \\
|c_{\mathrm{NL}}| S^{(2)} S^{(6)}
&\sim 1\times10^{-12}~
\mu\mathrm{K}^{-1} \cdot P_{\mathrm{sky}}^{\,2},
\end{align}
where
\begin{align}
P_{\mathrm{sky}}
=
\sqrt{
Q_{\mathrm{sky}}^{\,2}
+
U_{\mathrm{sky}}^{\,2}
+
V_{\mathrm{sky}}^{\,2}
}.
\end{align}

For a representative CMB temperature anisotropy of
$\Delta T \approx 30~\mu\mathrm{K}$ and
$P_{\mathrm{sky}} \approx 1~\mu\mathrm{K}$,
the nonlinear contributions are approximately
\begin{align}
|c_{\mathrm{NL}}| S^{(0)} S^{(4)}
&\sim 1\times10^{-7}~\mu\mathrm{K}, \\
|c_{\mathrm{NL}}| \left(S^{(2)}\right)^2
&\sim 4\times10^{-11}~\mu\mathrm{K}, \\
|c_{\mathrm{NL}}| S^{(2)} S^{(6)}
&\sim 1\times10^{-12}~\mu\mathrm{K}.
\end{align}

Thus, for the CMB anisotropy considered here, the additional
fourth-harmonic components generated by quadratic responsivity
non-linearity are negligible.
This estimate does not imply that detector non-linearity is generally
negligible, since coupling to larger signals, such as HWP-synchronous
signals or detector loading, can produce more significant effects
and may require calibration in a realistic time-ordered data model.

\subsection{Jones-matrix parameterization}
\label{subsec:jones_parameterization}
The optical response of CMB polarization experiments is often described using
Mueller matrices, particularly in studies of instrumental systematics and
polarization modulation \cite{ODea2007,Brown2009,LB_PTEP}. For non-depolarizing
optical elements, however, the Jones formalism provides an equivalent
description while allowing HWP non-idealities, such as differential
transmission, phase retardance, and cross-polarization, to be described directly
at the complex electric-field level. Jones-matrix descriptions have been used
to model HWP and instrumental non-idealities in CMB polarization experiments
\cite{Serena2022}. Following Giardiello et al.~\cite{Serena2022}, we adopt a
Jones-matrix description and extend it to include responses that depend on both
the HWP rotation angle and frequency.

We describe the RCWA-computed response as an effective Jones matrix depending on
frequency $\nu$ and HWP rotation angle $\rho$. 
We introduce a frequency-dependent matrix in the HWP frame
$\alpha(\nu)$ and express the jones matrix as
\begin{equation}
\boldsymbol{J}_{\alpha}(\nu,\rho)
=
\boldsymbol{R}^{\mathsf T}(\alpha(\nu))
\boldsymbol{R}^{\mathsf T}(\rho)
\boldsymbol{J}_{\rm RCWA}(\nu,\rho)
\boldsymbol{R}(\rho)
\boldsymbol{R}(\alpha(\nu)).
\label{eq:jones_alpha_basis}
\end{equation}
Here $\boldsymbol{R}$ denotes the two-dimensional rotation matrix. The angle
$\alpha(\nu)$ is defined from the principal axes of the rotation-averaged Jones
response.

In this basis, the response is written as a perturbation around an ideal HWP,
\begin{equation}
\boldsymbol{J}_{\alpha}(\nu,\rho)
=
\boldsymbol{J}_{\rm ideal}
+
\delta\boldsymbol{J}_{\alpha}(\nu,\rho),
\qquad
\boldsymbol{J}_{\rm ideal}
=
\begin{pmatrix}
1 & 0 \\
0 & -1
\end{pmatrix}.
\label{eq:jones_ideal_delta}
\end{equation}
The non-ideal part is expanded in Fourier modes of the HWP angle,
\begin{equation}
\delta\boldsymbol{J}_{\alpha}(\nu,\rho)
=
\sum_{m=0,\pm2,\pm4}
\boldsymbol{A}_{m}(\nu)e^{im\rho}.
\label{eq:jones_fourier_expansion}
\end{equation}
The DC component is parameterized in analogy with the Jones description of
Giardiello et al.~\cite{Serena2022},
\begin{equation}
\boldsymbol{A}_{0}(\nu)
=
\begin{pmatrix}
h_1(\nu) &
\zeta_1(\nu)e^{i\chi_1(\nu)}
\\
\zeta_2(\nu)e^{i\chi_2(\nu)} &
1-\{1+h_2(\nu)\}e^{i\beta(\nu)}
\end{pmatrix}.
\label{eq:jones_A0_parameterization}
\end{equation}
The parameters $h_1$ and $h_2$ describe deviations in the transmission along the
principal axes, $\beta$ represents the retardance error, and $\zeta_{1,2}$ and
$\chi_{1,2}$ describe the amplitudes and phases of the cross-polarization terms.
Now we introduce non-constant parameter with respect to $\rho$.
The additional $m=\pm2$ and $m=\pm4$ components capture the residual
rotation-angle dependence observed in the RCWA response. This provides a compact
extension of the conventional frequency-dependent Jones parameterization to an
effective response depending on both $\nu$ and $\rho$.


\begin{figure}
\centering
\includegraphics[width=1.0\linewidth]{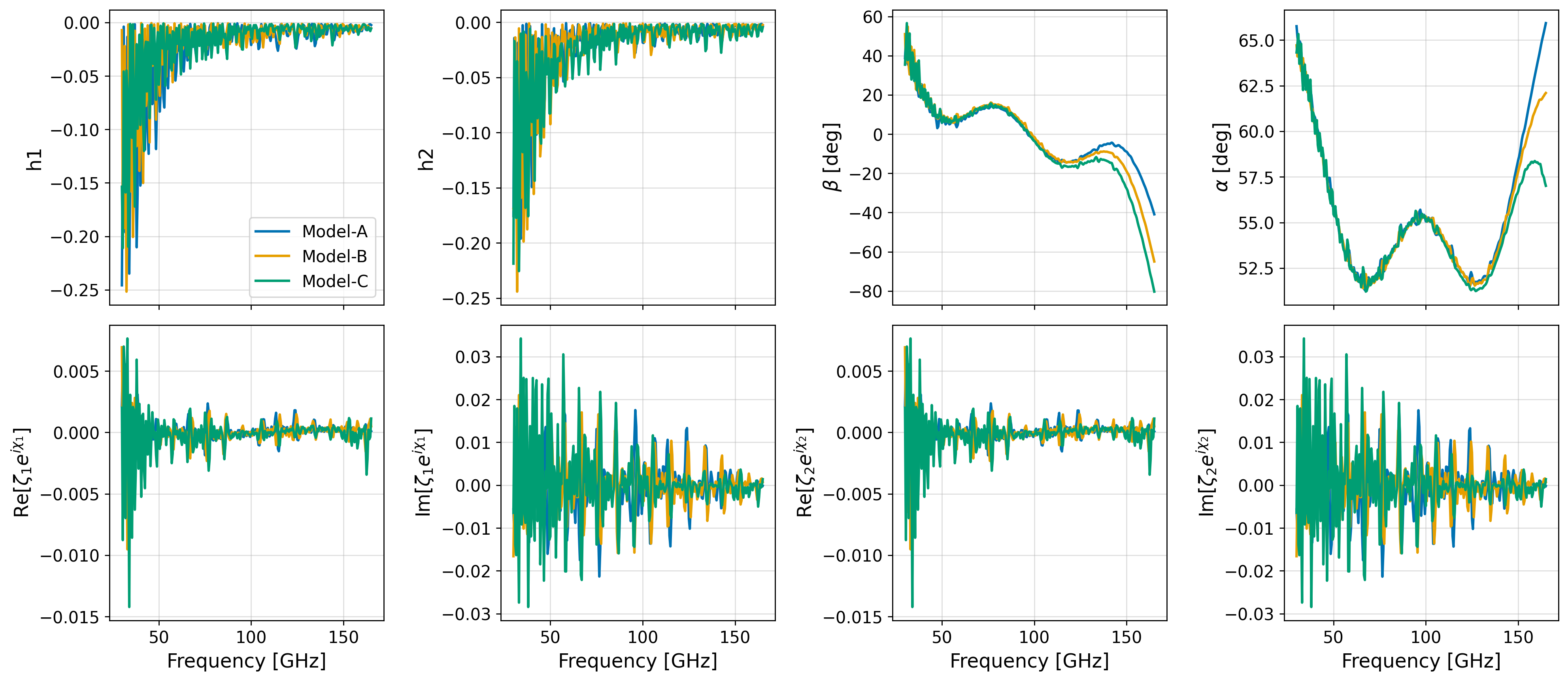}
\caption{
0th-ordered extracted Jones parameters for the three HWP models at $\Theta = 10^\circ$. The top row shows the
diagonal amplitude deviations $h_1$ and $h_2$, the retardance-error parameter
$\beta$, and the frequency-dependent principal-axis angle $\alpha$. The bottom
row shows the real and imaginary parts of the complex cross-polarization
coefficients $\zeta_1 e^{i\chi_1}$ and $\zeta_2 e^{i\chi_2}$. The
cross-polarization terms are shown in this complex form rather than as separate
amplitudes and phases in order to avoid ambiguities associated with phase
wrapping.
}
\label{fig:jones_parameters}
\end{figure}

\begin{figure}
    \centering
    \includegraphics[width=1.0\linewidth]{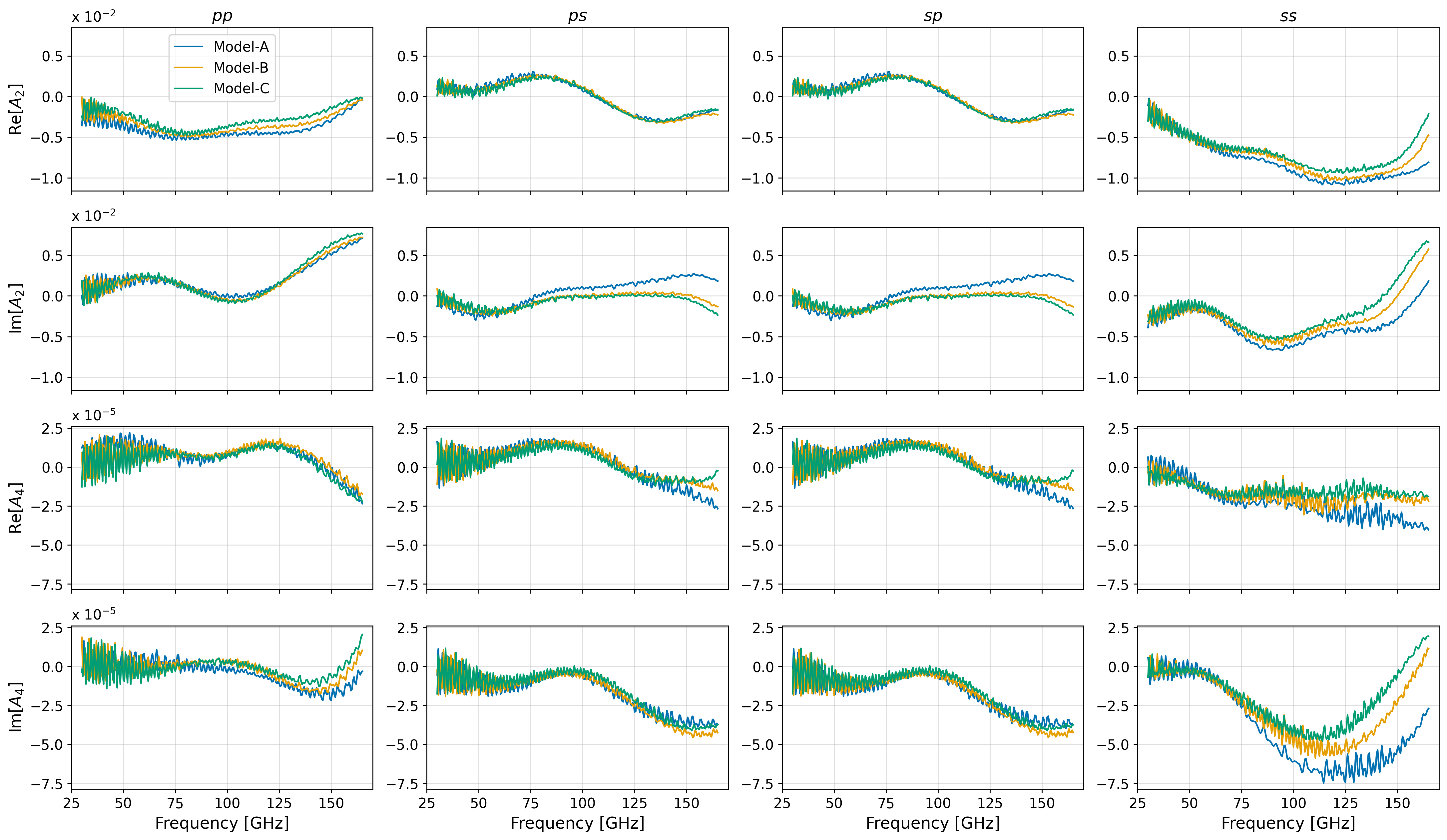}
    \caption{Fitted $2f$ and $4f$ Fourier components of the non-ideal Jones response as a function of frequency. Real and imaginary parts are shown for each Jones component and for the three SWS models.
}
    \label{fig:jones_2f4f}
\end{figure}

Figure~\ref{fig:jones_parameters} summarizes the extracted parameters for the
three HWP models at $\Theta = 10^\circ$. The diagonal terms $h_1$ and $h_2$ show the largest deviations
at the low-frequency end of the band and approach zero toward higher
frequencies, while the retardance-error parameter $\beta$ varies smoothly across
the band. The frequency-dependent angle matrix $\alpha$ captures the
rotation of the effective HWP axes after the geometrical HWP rotation has been
removed. The cross-polarization coefficients at 0th-order are small compared with the
diagonal response, but exhibit frequency-dependent structure that is retained in
the complex coefficients.
The residual HWP-angle dependence is shown in Fig.~\ref{fig:jones_2f4f}
through the fitted $2f$ and $4f$ Fourier coefficients. 
We plot the real and imaginary parts of $\boldsymbol{A}_{2}$ and $\boldsymbol{A}_{4}$ for each Jones component. 
The $2f$ coefficients are typically at the $10^{-3}$ level, whereas
the $4f$ coefficients are at the $10^{-5}$ level, showing that the dominant non-DC correction is carried by the $2f$, which term is prjected into $4f$ components in Mueller matrix.
The normalized RMS residual is below $5\times10^{-7}$ for all three models, supporting the validity of our formalism.

\section{Conclusion}
\label{sec:conclusion}

We modeled the instrumental polarization induced by a rotating multi-layer
sapphire HWP with SWS. The electromagnetic response was computed with RCWA as a
function of frequency, incident angle, and HWP rotation angle, and was converted
into Mueller-matrix coefficients for CMB polarimetry. By decomposing the Mueller
coefficients into harmonics of the HWP rotation angle, we evaluated the
HWP-synchronous components relevant to the detector signal.
For the three SWS models studied in this work, the instrumental polarization is in the fourth-harmonic Mueller coefficients.
Its amplitude is typically of order $10^{-4}$ in the CMB bands. This level is
expected to be treatable with the mitigation method proposed by Patanchon et
al.~\cite{Patanchon_2024}, provided that the HWP-induced signal is characterized and included in the
time-ordered data model. The unintended conversion between linear polarizations
is much smaller. The corresponding additional polarization-angle rotation is
approximately 30 arcsec at maximum.

The foreground spectral indices within representative ranges changes the
band-averaged efficiency only at the percent level or below for most frequency
bands. The effect of the difference with the band coverage is also negligible if the instrumental
polarization does not need to be modeled at better than the percent level. The
simple estimate of detector non-linearity gives a negligible contribution for
CMB anisotropy signals, although larger loading terms may still need to be
included in a realistic detector model.

The comparison among the SWS models shows that the size of the instrumental
polarization is not mainly determined by the asymmetry model itself. By comparing with the SWS model with realistic model, the higher transmittance represented by $M_{II}^{(0)}$ can lead to lower instrumental polarization. This can imply that the SWS design should be optimized not only for retardance but also for improving transmission. 
Model-A remains attractive because it can reduce the required
sapphire thickness while keeping comparable optical performance, which is useful
for mass-limited satellite missions.

\appendix
\section{Modeling SWS}
\label{appendix:SWS}

We model asymmetric AR-SWS using the ellipse contour along the $z$ -axis. Taking into account the variables height $h$, pitch $p_x,~p_y$, top radius $r_{xt},~r_{yt}$ and bottom radius $r_{xb},~r_{yb}$, we create the 3D model of AR SWS as
\begin{align}
    \frac{\tilde{x}^{2}}{a(z)^{2}} &+ \frac{\tilde{y}^{2}}{b(z)^{2}} = 1\\
    r_x(z) &= r_{xt} + (r_{xb} - r_{xt})\left(1 - (z/h)^{\alpha_x}\right) \\
    r_y(z) &= r_{yt} + (r_{yb} - r_{xb})\left(1 - (z/h)^{\alpha_y}\right),
\end{align}
where
\begin{align}
\label{eq:conditionx}
    \tilde{x} &=
        \begin{cases}
            -\dfrac{p_x}{2}, & x < -\dfrac{p_x}{2}, \\
            x, & -\dfrac{p_x}{2} \le x \le \dfrac{p_x}{2}, \\
            \dfrac{p_x}{2}, & x > \dfrac{p_x}{2},
        \end{cases}
\end{align}

\begin{align}
\label{eq:conditiony}
    \tilde{y} &=
        \begin{cases}
            -\dfrac{p_y}{2}, & y < -\dfrac{p_y}{2}, \\
            y, & -\dfrac{p_y}{2} \le y \le \dfrac{p_y}{2}, \\
            \dfrac{p_y}{2}, & y > \dfrac{p_y}{2},
        \end{cases}
\end{align}
The bottom radius can exceed $p_x$ and $p_y$ with condition as shown in Equation~\ref{eq:conditionx} and \ref{eq:conditiony}, and the curvature of the structure is defined with the bottom radius and $\alpha_x$ and $\alpha_y$, introduced by R. Takaku et al.~\cite{Takaku2020} to evaluate the AR performance of the shape curvature.
This model represents the actual structures fabricated with ultra-short pulsed laser ablation. This model makes the different height in the cross-section of $x$, $y$ and cut in the corner. We define them as saddle heights $h_x,~h_y$ and total height $h_t$, respectively. To fare comparison, we keep the same total height between different types of asymmetry, we also define the following condition:
\begin{align}
\label{eq:condition_for_total_height}
    \frac{(p_x/2)^2}{r_{xb}^2} &+ \frac{(p_y/2)^2}{r_{yb}^2} = 1.
\end{align}
From this condition, given $r_{xb}$ we derive the bottom radius in $y$. 

\section{Conversion between linear polarization}
\label{AppendixB}

The polarization angle on the sky, $\alpha$, is defined as an
angle between the electric field vector on the sky and $X_{\mathrm{sky}}$
axis in Figure~2. Using $\alpha$, we can express the Stokes
parameters like $Q_{\mathrm{sky}} = P \cos 2\alpha$ and
$U_{\mathrm{sky}} = P \sin 2\alpha$, where $0 < P \leq I_{\mathrm{sky}}$.
When a HWP works as shown in Equation~(3), the detected power is given by
\begin{align}
2S
=
I_{\mathrm{sky}}
+
P \cos\left(4\rho - 4\Phi + 4\chi - 2\psi - 2\alpha\right).
\end{align}

The matrix varies slightly from the ideal case and some
of the elements result in
\begin{align}
M'_{QQ}
&=
M_{QQ}\cos 4(\rho-\Phi), \\
M'_{QU}
&=
M_{QQ}(1+\Delta_{QU})\sin 4(\rho-\Phi+\delta_{QU}) \\
&\sim
M_{QQ}\left[(1+\Delta_{QU})\sin 4(\rho-\Phi)
+ \delta_{QU}\cos 4(\rho-\Phi)\right], \\
M'_{UQ}
&=
M_{QQ}(1+\Delta_{UQ})\sin 4(\rho-\Phi+\delta_{UQ}) \\
&\sim
M_{QQ}\left[(1+\Delta_{UQ})\sin 4(\rho-\Phi)
+ \delta_{UQ}\cos 4(\rho-\Phi)\right], \\
M'_{UU}
&=
-M_{QQ}(1+\Delta_{UU})\cos 4(\rho-\Phi+\delta_{UU}) \\
&\sim
-M_{QQ}\left[(1+\Delta_{UU})\cos 4(\rho-\Phi)
- \delta_{UU}\sin 4(\rho-\Phi)\right].
\end{align}

Note that $|\Delta_{XY}| \ll 1$ and $|\delta_{XY}| \ll 1$.
After lengthy calculation, the detected signal is expressed as
\begin{align}
2S'
&\sim
I_{\mathrm{sky}}
\nonumber\\
&\quad
+ P M_{QQ}\mathcal{A}
\cos\left(4\rho - 4\Phi + 4\chi - 2\psi - 2\alpha - 2\Delta\alpha\right),
\end{align}
where
\begin{align}
\mathcal{A}
&\sim
1
+
\left[\Delta_{UU}\sin(2\psi-2\chi)\sin(2\alpha-2\chi)\right.
\nonumber\\
&\quad
- \delta_{QU}\cos(2\psi-2\chi)\sin(2\alpha-2\chi)
\nonumber\\
&\quad
- \delta_{UQ}\sin(2\psi-2\chi)\cos(2\alpha-2\chi)\cos(2\psi+2\alpha-4\chi)
\nonumber\\
&\quad
+ \Delta_{QU}\cos(2\psi-2\chi)\sin(2\alpha-2\chi)
\nonumber\\
&\quad
+ \Delta_{UQ}\sin(2\psi-2\chi)\cos(2\alpha-2\chi)
\nonumber\\
&\quad
\left.
+ \delta_{UU}\sin(2\psi-2\chi)\sin(2\alpha-2\chi)\right]\sin(2\psi+2\alpha-4\chi),
\end{align}
\begin{align}
2\Delta\alpha
&\sim
\frac{\delta_{QU}+\delta_{UQ}+\delta_{UU}}{4}
\nonumber\\
&\quad
+
\frac{\Delta_{QU}-\Delta_{UQ}+\Delta_{UU}}{4}\sin(4\alpha-4\chi)
\nonumber\\
&\quad
+
\frac{\delta_{QU}-\delta_{UQ}+\delta_{UU}}{4}\cos(4\alpha-4\chi)
\nonumber\\
&\quad
+
\frac{\cos(4\psi-4\chi)}{4}
\Big[
-(\delta_{QU}-\delta_{UQ}-\delta_{UU})
\nonumber\\
&\qquad\qquad\qquad
+ (\Delta_{QU}+\Delta_{UQ}-\Delta_{UU})\sin(4\alpha-4\chi)
\nonumber\\
&\qquad\qquad\qquad
+ (\delta_{QU}+\delta_{UQ}-\delta_{UU})\cos(4\alpha-4\chi)
\Big]
\nonumber\\
&\quad
+
\frac{\sin(4\psi-4\chi)}{4}
\Big[
-(\Delta_{QU}-\Delta_{UQ}-\Delta_{UU})
\nonumber\\
&\qquad\qquad\qquad
+ (\delta_{QU}+\delta_{UQ}-\delta_{UU})\sin(4\alpha-4\chi)
\nonumber\\
&\qquad\qquad\qquad
+ (\Delta_{QU}+\Delta_{UQ}-\Delta_{UU})\cos(4\alpha-4\chi)
\Big].
\end{align}

When the angles $\alpha$ and $\psi$ satisfy the relation,
\begin{align}
\tan(4\psi-4\chi)
&=
\frac{
\begin{aligned}
&-\left(\Delta_{QU}-\Delta_{UQ}-\Delta_{UU}\right) \\
&+\left(\delta_{QU}+\delta_{UQ}-\delta_{UU}\right)\sin(4\alpha-4\chi) \\
&+\left(\Delta_{QU}+\Delta_{UQ}-\Delta_{UU}\right)\cos(4\alpha-4\chi)
\end{aligned}
}{
\begin{aligned}
&-\left(\delta_{QU}-\delta_{UQ}-\delta_{UU}\right) \\
&+\left(\Delta_{QU}+\Delta_{UQ}-\Delta_{UU}\right)\sin(4\alpha-4\chi) \\
&+\left(\delta_{QU}+\delta_{UQ}-\delta_{UU}\right)\cos(4\alpha-4\chi)
\end{aligned}
}.
\end{align}
the angular shift $2\Delta\alpha$ becomes a maximum or minimum.


\acknowledgments
This work was supported by JSPS KAKENHI Grant Number JP24K17064.
This work has also received funding from the European Union's Horizon 2020 research and innovation programme under grant agreement No.~101007633 CMB-Inflate.

RT and GP acknowledge the RECTOR program in Okayama University.



\bibliographystyle{unsrt}
\bibliography{bib}







\end{document}